\documentclass[12pt,a4j]{article}
\usepackage[dvips]{graphicx}
\usepackage{amsmath}
\usepackage{amssymb}
\usepackage{enumerate}
\usepackage{amscd}

\begin{document}
\baselineskip=6
mm
\centerline{\bf Multisoliton solutions of the two-component Camassa-Holm system}\par
\centerline{\bf and their reductions}\par
\bigskip
\centerline{Yoshimasa Matsuno\footnote{{\it E-mail address}: matsuno@yamaguchi-u.ac.jp}}\par
\centerline{\it Division of Applied Mathematical Science, } \par
\centerline{\it Graduate School of Sciences and Technology for Innovation,}
\centerline{\it Yamaguchi University, Ube, Yamaguchi 755-8611, Japan} \par
\bigskip
\bigskip
\bigskip
\noindent {\bf Abstract} \par
\noindent We develop a systematic procedure for constructing soliton solutions of an integrable two-component
Camassa-Holm (CH2) system. The parametric representation of the multisoliton solutions is obtained
by using a direct method combined with a reciprocal transformation. The properties of the
solutions are then investigated in detail focusing mainly on the smooth one- and two-soliton solutions.
The general $N$-soliton case is described shortly. Subsequently, we show that the CH2  system reduces to the
CH equation and the two-component Hunter-Saxton (HS2) system by means of appropriate limiting procedures.
The corresponding expressions of the multisoliton solutions are presented in  parametric forms, reproducing
the existing results for the reduced equations. Last, we discuss  the reduction from the HS2 system to the HS equation.
\par

\newpage
\leftline{\bf  1. Introduction} \par
\bigskip
\noindent  In this paper, we consider the following two-component generalization of the  Camassa-Holm (CH) equation 
$$m_t+um_x+2mu_x+\rho\rho_x=0, \eqno(1.1a)$$
$$\rho_t+(\rho u)_x=0, \eqno(1.1b)$$
which is abbreviated as the CH2 system.
Here, $u=u(x,t), \rho=\rho(x,t)$ and $m=m(x,t)\equiv u-u_{xx}+\kappa^2$ are real-valued functions of time $t$ and a spatial variable $x$, and 
the subscripts $x$ and $t$ appended to $u$ and $\rho$ denote partial differentiation. The parameter $\kappa$ in the expression of $m$ is assumed to be
a non-negative real number. \par
The CH2 system (1.1) has been derived for the first time in [1] in search of
the bi-Hamiltonian formulation of integrable nonlinear evolution equations. Actually,  the  system can be represented  as the dual bi-Hamiltonian
system for a coupled Korteweg-de Vries equation introduced independently  by Zakharov [2] and Ito [3]. Later, a similar system with the coefficient of $\rho\rho_x$ in (1.1a) being minus was
studied [4-6]. In particular,  a reciprocal transformation between the system and the first negative flow of the AKNS hierarchy was established in [6].
In the physical context, on the other hand, the  CH2 system with $\kappa=0$ was  derived  by applying  an asymptotic analysis to  the fully nonlinear Green-Naghdi equations 
for shallow water waves, where $u$ represents 
 the horizontal velocity and  $\rho$ is related to the depth of the fluid in the first approximation [7]. The same system with $\kappa\not= 0$ was also  obtained from the basic Euler
system for an incompressible fluid with a constant vorticity [8]. 
One can also consult Ref. [9]  as for a brief history of the CH2 system.
\par
One remarkable feature of the CH2 system is that it is a completely integrable system. Indeed, it has  a Lax representation given by 
$$\Psi_{xx}=\left(-\lambda^2\rho^2+\lambda m+{1\over 4}\right)\Psi, \eqno(1.2a)$$
$$\Psi_t=\left({1\over 2\lambda}-u\right)\Psi_x+{u_x\over 2}\Psi, \eqno(1.2b)$$
where $\lambda$ is  the spectral parameter [7, 8].  It turns out that the compatibility condition of the linear system (1.2) yields (1.1), thus enabling us to apply the
inverse scattering transform method (IST) [10, 11].
A number of works have been devoted to the study of the mathematical properties of (1.1). 
For example, some conditions were provided for the wave breaking and the existence of the traveling waves [7, 12, 13]. The explicit solitary wave solutions were obtained
by using the method of dynamical systems [14, 15], and the general multisoliton solutions were constructed by means of the IST [16]. 
More precisely, the IST is reformulated as a Riemann-Hilbert problem [11], and the $N$-soliton solution is given by a parametric form.
However, the analysis of multisoliton solutions has not been done as yet.
\par
Various reductions  are possible for  the CH2 system while preserving its integrability.  Specificallly, the reduction to the
CH equation is of great importance. This can be accomplished simply by putting $\rho=0$ in (1.1), giving [17]
$$u_t+2\kappa^2u_x-u_{xxt}+3uu_x=2u_xu_{xx}+uu_{xxx}. \eqno(1.3)$$
The CH equation describes the unidirectional propagation of shallow water waves over a flat bottom. Its structure has been studied extensively
from both theoretical and numerical points of view [18, 19].
The Lax representation associated with the CH equation can be obtained simply by putting $\rho=0$ in (1.2). This enables us to apply the IST which
has been successfully used  for various integrable soliton equations such as the Korteweg-de Vries (KdV) and nonlinear Schr\"odinger equations.
Unlike the KdV equation which is a typical model of shallow water waves, the CH equation could explain the wave breaking as well as the existence of peaked waves (or peakons)
which are inherent in the basic Euler system. \par
Another reduction is the two-component Hunter-Saxton (HS2) system which can be derived by means of the short-wave limit of the CH2 system. It has the same form as 
the system (1.1) with the variable $m$ replaced by $-u_{xx}+\kappa^2$.  Explicitly, it can be written in the form
$$u_{xxt}-2\kappa^2 u_x+uu_{xxx}+2u_xu_{xx}-\rho\rho_x=0, \quad \rho_t+(\rho u)_x=0. \eqno(1.4)$$
 Furthermore, on taking $\rho=0$, the HS2 system (1.4) reduces to 
 $$u_{xxt}-2\kappa^2u_x+uu_{xxx}+2u_xu_{xx}=0. \eqno(1.5)$$
 In the case of $\kappa=0$, equation (1.5) becomes
  the classical Hunter-Saxton (HS) equation  
 which is a model for describing the propagation of weakly nonlinear orientation waves in a massive nematic liquid crystal director field [20].
 We refer to (1.5) as the HS equation hereafter. 
\par
The purpose of the present paper is to develop a systematic method for obtaining the multisoliton solutions of the CH2 system  and investigate their properties. 
Subsequently, a reduction procedure is performed to obtain the multisoliton solutions of the CH equation and the HS2 system from those of the CH2 system. We impose the boundary conditions
$u(x,t) \rightarrow 0$ and $\rho(x,t) \rightarrow \rho_0$ as $|x| \rightarrow \infty$, where $\rho_0$ is a positive constant. These boundary conditions are 
consistent with the hydrodynamic derivation of the system [7, 8]. 
A direct method is employed to obtain solutions which worked effectively for
the construction of the soliton solutions of 
 the CH equation [21] and  the modified  CH equations [22, 23].  \par
This paper is organized as follows. In section 2, we transform the CH2 system to a
 system of partial differential equations (PDEs) by means of a reciprocal transformation similar to that employed for the CH and modified CH equations [21-23].
 We then perform the bilinearization of the latter system through appropriate dependent variable transformations.  Following the standard procedure of the
 bilinear transformation method [24, 25], we construct the $N$-soliton solution of the bilinear equations in terms of the tau-functions, where $N$ is an arbitrary positive integer, thus
 obtaining the parametric representation for the $N$-soliton solution of the system (1.1). 
 The dispersion relation of the soliton is explored in detail to feature its propagation characteristics. 
 In section 3, we investigate the properties of the soliton solutions.
 First, we address the one-soliton solutions, showing that the profile of $\rho$  always takes the form of bright soliton whereas 
 that of $u$ takes both bright and dark solitons depending on the dispersion relation of the soliton.
 Subsequently, the asymptotic analysis of the $N$-soliton solution is performed to derive the formula for the phase shift. Last, the interaction process
 of two solitons is exemplified for both overtaking and head-on collisions. 
 In section 4,  we carry out various reductions of the CH2 system. Specifically, by
 introducing appropriate scaling variables, we demonstrate that the CH2 system reduces to the CH equation in the limit $\rho_0 \rightarrow 0$, and
 recover the $N$-soliton solution of the CH equation as well as the formula for the phase shift. We also show that the short-wave limit of the CH2 system leads to the
 HS2 system, and the $N$-soliton solution of the latter system is recovered from that of the former system. Then, we give a brief summary about the reduction to the HS equation.
  Section 5 is devoted to some concluding
 remarks.  In appendix A, we detail the bilinearization of the CH2 system.
 In appendix B, we provide a proof of  the bilinear identities for the tau-functions associated with the $N$-soliton solution of the CH2 system. \par
 \bigskip
 \noindent{\bf 2. Exact method of solution} \par
 \bigskip
 \noindent In this section, we develop a systematic method for constructing the multisoliton solutions of the CH2 system. To this end, we
 employ an exact method of solution which is referred to as the direct method [24] or the bilinear transformation method [25].  When compared with the IST, this method is
 an especially powerful technique for  obtaining particular solutions like soliton and periodic wave solutions. After transforming the system (1.1) to an equivalent system
 of PDEs by a reciprocal transformation, we bilinearize the latter system and then solve it in terms of the tau-functions, thus giving rise to the parametric representation of the
 $N$-soliton solution. \par
 \bigskip
 \noindent{\it 2.1. Reciprocal transformation}\par
 \bigskip
 \noindent First of all, we introduce the reciprocal transformation $(x ,t) \rightarrow (y, \tau)$ according to 
 $$dy=\rho\,dx-\rho u\,dt,\qquad d\tau=dt. \eqno(2.1a)$$
 Then, the $x$ and $t$ derivatives transform  as
 $${\partial\over\partial x}=\rho{\partial\over\partial y}, \qquad {\partial\over\partial t}
 ={\partial\over\partial \tau}-\rho u{\partial\over\partial y}. \eqno(2.1b)$$
 Applying the transformation (2.1) to the system (1.1), we obtain the system of PDEs 
 $$\left({m\over \rho^2}\right)_\tau+\rho_y=0, \eqno(2.2a)$$
 $$\rho_\tau+\rho^2u_y=0. \eqno(2.2b)$$
 It then follows from $(2.1b)$ that the variable $x=x(y, \tau)$ obeys a system of linear PDEs
$$x_y={1\over \rho}, \eqno(2.3a)$$
$$ x_\tau=u. \eqno(2.3b)$$
The system of equations (2.3) is integrable since its compatibility condition $x_{\tau y}=x_{y\tau}$ is assured by virtue of
$(2.2b)$. \par
  Now, the quantity $m=u-u_{xx}+\kappa^2$ in (1.1) can be rewritten in  terms of the new coordinate system as
 $$m=u+\rho({\rm ln}\,\rho)_{\tau y}+\kappa^2,\eqno(2.4)$$
 where we have used $(2.2b)$ to replace $u_y$ by $-\rho_\tau/\rho^2$. Let us introduce the new dependent variable $Y=Y(y, \tau)$ by the relation
 $${m\over \rho^2}-{\kappa^2\over \rho_0^2}=Y_y. \eqno(2.5)$$
 Subsituting (2.5) into $(2.2a)$ and then integrating the resultant expression by $y$ 
 under the boundary conditions $Y_\tau\rightarrow 0$ and $\rho\rightarrow \rho_0$ as $|y|\rightarrow \infty$, we obtain
 $$\rho=\rho_0-Y_\tau. \eqno(2.6)$$
 \par
 The following proposition is the starting point in the present analysis.\par
 \bigskip
 \noindent {\bf Proposition 2.1.} {\it The variables $x$ and $Y$ satisfy the system of PDEs
  $$x_y(\rho_0-Y_\tau)=1,    \eqno(2.7)$$
  $$(\rho_0-Y_\tau)\left({\kappa^2\over \rho_0^2}+Y_y\right)=x_\tau x_y-[(\rho_0-Y_\tau)x_{\tau y}]_y+\kappa^2 x_y. \eqno(2.8)$$}
  \par
  \noindent {\bf Proof.}\ Equation (2.7) follows immediately from $(2.3a)$ and (2.6).
 If we substitute $m$ from (2.5) into (2.4) and use $(2.3a)$ and $(2.3b)$ to express $\rho$ and $u$ in terms of $x_y$ and $x_\tau$, respectively,  (2.4) becomes
 $${\kappa^2\over \rho_0^2}+Y_y=x_\tau x_y^2-x_{\tau yy}+{x_{\tau y}x_{yy}\over x_y}+\kappa^2x_y^2. $$
 Dividing this expression by $x_y$ and using (2.7), we arrive at (2.8). \hspace{\fill} $\square$ \par
 \bigskip
 \noindent{\it 2.2. Bilinearization}\par
 \bigskip
\noindent In applying the bilinear transformation method to the given nonlinear equations, the first step is to transform the
 equations into the bilinear equations, which we shall now demonstrate.
 To this end, we introduce the dependent variable transformations
 $$x={y\over \rho_0}+{\rm ln}\,{\tilde f\over f}+d, \eqno(2.9)$$
 $$Y={\rm i}\,{\rm ln}\,{\tilde g\over g}, \eqno(2.10)$$
 where $f, \tilde f, g$ and $\tilde g$ are tau-functions and $d$ is an arbitrary constant. 
  One advantage of  the form (2.10) 
 is that the structure of the system of bilinear equations becomes transparent when compapred with  the introduction of
  another form like $Y=2\,{\tan}^{-1}({\rm Im}\,g/{\rm Re}\,g)$. This facilitates the analysis, in particular the construction of solutions.
 Obviously,  the definition of $Y$ from (2.5) implies that it can be taken as a real quantity
 which is achieved simply if one chooses the tau-function $\tilde g$ as a complex conjugate of $g$. 
 This recipe can be used successfully in constructing real soliton solutions, as will be manifested in theorem 2.2.
 \par
  Now, we establish the following proposition. \par
 \bigskip
 \noindent {\bf Proposition 2.2.}\ {\it Consider the following system of bilinear equations for $f, \tilde f, g$ and $\tilde g$:
 $$D_y\tilde f\cdot f+{1\over \rho_0}(\tilde f f-\tilde g g)=0, \eqno(2.11)$$
 $${\rm i}D_\tau\tilde g\cdot g+ \rho_0(\tilde f f-\tilde g g)=0, \eqno(2.12)$$
 $$D_\tau D_y\tilde f\cdot f +{1\over \rho_0}D_\tau\tilde f\cdot f+\kappa^2D_y\tilde f\cdot f=0, \eqno(2.13)$$
 $$D_\tau D_y\tilde g\cdot g-{\rm i}\,{\kappa^2\over \rho_0^2}\,D_\tau\tilde g\cdot g+{\rm i}\rho_0 D_y\tilde g\cdot g=0, \eqno(2.14)$$
 where the bilinear operators  are defined by
$$D_y^mD_\tau ^nf\cdot g=\left(\partial_y-\partial_{y^\prime} \right)^m\left(\partial_\tau-\partial_{\tau^\prime} \right)^n
f(y,\tau)g(y^\prime,\tau^\prime)|_{y^\prime=y,\,\tau^\prime=\tau}, \qquad (m, n=0, 1, 2, ...).   \eqno(2.15)$$
Then, the solutions of this system of equations solve the equations (2.7) and (2.8). } \par
The proof of proposition 2.2 will be detailed in appendix A. \par

\bigskip
\noindent{\it 2.3. Parametric representations of the solutions}\par
\bigskip
\noindent {\bf Theorem 2.1.} {\it  The two-component CH system (1.1) admits the parametric representations of the solutions
$$u(y, \tau)=\left({\rm ln}\,{\tilde f\over f}\right)_\tau, \eqno(2.16)$$
$$\rho(y, \tau)=\rho_0-{\rm i}\left({\rm ln}\,{\tilde g\over g}\right)_\tau, \eqno(2.17)$$
 $$x(y, \tau)={y\over \rho_0}+{\rm ln}\,{\tilde f\over f}+d. \eqno(2.18)$$
 }\par
\noindent {\bf Proof.} The expression (2.16) follows by introducing $(2.9)$ into $(2.3b)$ whereas the expression (2.17)
comes from (2.6) and (2.10). The expression (2.18) is just (2.9). \hspace{\fill} $\square$ \par
\bigskip
\noindent {\bf Remark 2.1.}\ The parametric representations of $1/\rho$ and $m/\rho^2$ in terms of the tau-functions
are also available from $(2.3a)$, (2.5), (2.9) and (2.10). Explicitly, they read
$${1\over \rho}={1\over \rho_0}+\left({\rm ln}\,{\tilde f\over f}\right)_y, \eqno(2.19)$$
$${m\over \rho^2}={\kappa^2\over \rho_0^2}+{\rm i}\left({\rm ln}\,{\tilde g\over g}\right)_y. \eqno(2.20)$$
\par
\bigskip
\noindent{\it 2.4. $N$-soliton solution}\par
\bigskip
\noindent {\bf Theorem 2.2.} {\it The tau-functions $f, \tilde f, g$ and $\tilde g$ constituting the $N$-soliton solution of the system of bilinear equations 
(2.11)-(2.14) are given by the expressions
$$f=\sum_{\mu=0,1}{\rm exp}\left[\sum_{j=1}^N\mu_j\left(\xi_j+\phi_j\right)
+\sum_{1\le j<l\le N}\mu_j\mu_l\gamma_{jl}\right], \eqno(2.21a)$$
$$\tilde f=\sum_{\mu=0,1}{\rm exp}\left[\sum_{j=1}^N\mu_j\left(\xi_j-\phi_j\right)
+\sum_{1\le j<l\le N}\mu_j\mu_l\gamma_{jl}\right], \eqno(2.21b)$$
  $$g=\sum_{\mu=0,1}{\rm exp}\left[\sum_{j=1}^N\mu_j\left(\xi_j+{\rm i}\psi_j\right)
+\sum_{1\le j<l\le N}\mu_j\mu_l\gamma_{jl}\right], \eqno(2.22a)$$
 $$\tilde g=\sum_{\mu=0,1}{\rm exp}\left[\sum_{j=1}^N\mu_j\left(\xi_j-{\rm i}\psi_j\right)
+\sum_{1\le j<l\le N}\mu_j\mu_l\gamma_{jl}\right], \eqno(2.22b)$$
where
$$\xi_j=k_j\left(y-c_j\tau-y_{j0}\right), \qquad (j=1, 2, ..., N),\eqno(2.23a)$$
$${\rm  e}^{\gamma_{jl}}={\kappa^2(c_j-c_l)^2-\rho_0(k_j-k_l)c_jc_l(c_jk_j-c_lk_l)\over \kappa^2(c_j-c_l)^2-\rho_0(k_j+k_l)c_jc_l(c_jk_j+c_lk_l)}, \qquad (j, l=1, 2, ..., N; j\not=l),
\eqno(2.23b)$$
$${\rm e}^{-\phi_j}=\sqrt{(1-\rho_0k_j)c_j-\rho_0\kappa^2\over (1+\rho_0k_j)c_j-\rho_0\kappa^2},\qquad (j=1, 2, ..., N),\eqno(2.23c)$$
$${\rm e}^{-{\rm i}\psi_j}=\sqrt{\left({\kappa^2\over \rho_0}-{\rm i}\rho_0k_j\right)c_j+\rho_0^2\over \left({\kappa^2\over \rho_0}+{\rm i}\rho_0k_j\right)c_j+\rho_0^2},\qquad (j=1, 2, ..., N),\eqno(2.23d)$$
and $c_j$ is the velocity of $j$th soliton in the $(y, \tau)$ coordinate system which is given by  the solution of the quadratic equation
$$(1-\rho_0^2k_j^2)c_j^2-2\rho_0\kappa^2c_j-\rho_0^4=0, \qquad (j=1, 2, ..., N).\eqno(2.23e)$$
Here, $k_j$ and $y_{j0}$ are arbitrary complex parameters satisfying the conditions $k_j\not= k_l$
for $j\not=l$.  The notation $\sum_{\mu=0,1}$
implies the summation over all possible combinations of $\mu_1=0, 1, \mu_2=0, 1, ..., 
\mu_N=0, 1$}.\par 
\bigskip
A proof of theorem 2.2 will be given in appendix B in which the tau-functions (2.21) and (2.22) are shown to  satisfy the system of bilinear equations (2.11)-(2.14) by
means of mathematical induction. \par
\bigskip
\noindent {\bf Remark 2.2.} The bilinear equations (2.13) and (2.14) arise from the reduction of the BKP family of  integrable soliton equations [26, 27].
The tau-functions associated with the $N$-soliton solutions of these equations have the same forms as those given by (2.21) and (2.22). Within this framework, however, 
the parameters $c_j$ and $k_j$ in $(2.23b)$ can be taken independently. On the other hand,  for the present $N$-soliton solutions, both parameters are related to each other by the quadratic equation  $(2.23e)$.
This follows from the requirement that the tau-functions solve the bilinear equations (2.11) and (2.12) simultaneously.
\par
\bigskip
The parametric representation of the $N$-soliton solution given by (2.16)-(2.18) with the tau-functions (2.21) and (2.22) is characterized by the $2N$ complex parameters $k_j$ and $y_{j0}\ (j=1, 2. ..., N)$. 
The parameters $k_j$ determine the amplitude and the velocity of the  solitons, whereas
the parameters $y_{j0}$ determine the position (or phase) of the solitons. If we impose the conditions $\tilde f=f^*$ and $\tilde g=g^*$ where the asterisk denotes complex conjugate, then the solutions
become real functions of $x$ and $t$. Note, however that they would yield multi-valued functions unless certain conditions are imposed on the parameters $k_j (j=1, 2, .., N)$. 
The similar  situation has already been encountered in investigating the structure of the soliton solutions of the CH and modified CH equations [21-23]. 
We will address this issue in the next section where the detailed analysis of the
soliton solutions will be performed. \par

\begin{figure}[t]
\begin{center}
\includegraphics[width=8cm]{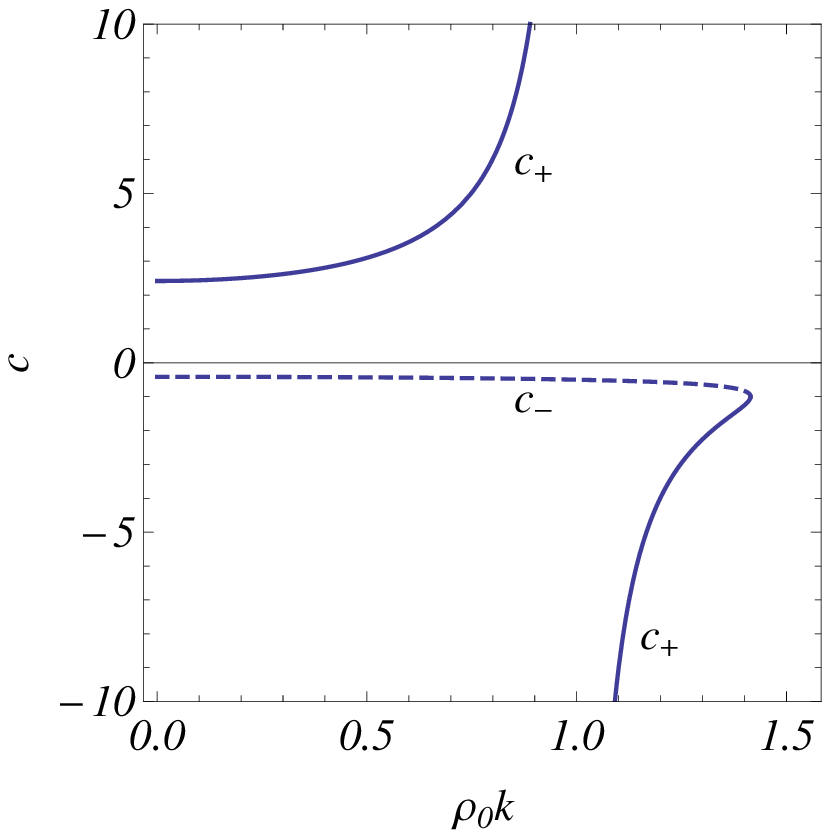}
\end{center}
\noindent {{\bf Figure 1.}\ The velocity $c=c_\pm$ of the soliton as a function of $\rho_0 k$ for $\rho_0=1$ and $\kappa=1$: $c_+$(solid curve), $c_-$(dashed curve).}
\end{figure}

Before proceeding, we investigate the characteristics of the velocity of the soliton in the $(y, \tau)$ coordinate system. 
As will be discussed in section 3.1, the corresponding velocity in $(x, t)$ coordinate system is given simply by $c_j/\rho_0$. 
The quadratic equation $(2.23e)$ has two roots
$$c_j={\rho_0\over 1-(\rho_0k_j)^2}\,(\kappa^2+d_j)={\rho_0^3\over d_j-\kappa^2}, \qquad (j=1, 2, ..., N),\eqno(2.24a)$$
where
$$d_j=\epsilon_j\sqrt{\kappa^4+\rho_0^2-\rho_0^4k_j^2},\qquad (\epsilon_j=\pm1, \quad j=1, 2, ..., N). \eqno(2.24b)$$
To assure the reality of $c_j$, one must impose the condition for the parameter $\rho_0 k_j$. Actually, it must lie in the interval
$$0<\rho_0 k_j<\sqrt{\kappa^4+\rho_0^2}/ \rho_0,\qquad (j=1, 2, ..., N), \eqno(2.25)$$
where we have assumed $k_j>0\ (j=1, 2, ..., N)$.
Figure 1 plots the velocities $c_+\equiv c_j(\epsilon_j=+1)$ and  $c_-\equiv c_j(\epsilon_j=-1)$ as a function of $\rho_0 k\equiv \rho_0 k_j$.
The velocity $c_+$ is positive for $0<\rho_0 k<1$ and negative for $1<\rho_0 k<\sqrt{\kappa^4+\rho_0^2}/\rho_0$. It exhibits the singularity at $\rho_0 k=1$.
Specifically,
$$\rho_0\left(\kappa^2+\sqrt{\kappa^4+\rho_0^2}\right)<c_+ <\infty, \ (0<\rho_0k<1), \eqno(2.26a)$$
$$-\infty<c_+ <-\rho_0^3/ \kappa^2,\ \left(1<\rho_0k<{\sqrt{\kappa^4+\rho_0^2}/ \rho_0}\right). \eqno(2.26b)$$
On the other hand, the velocity $c_-$ is a continuous function of $\rho_0 k$ and takes negative values in the interval (2.25), as indicated by the inequality
$$-\rho_0^3/\kappa^2<c_-<-\rho_0\left(\sqrt{\kappa^4+\rho_0^2}-\kappa^2\right),\ \left(0<\rho_0k<{\sqrt{\kappa^4+\rho_0^2}/ \rho_0}\right). \eqno(2.27)$$
In particular, $c_-=-\rho_0^3/(2\kappa^2)$ at $\rho_0k=1$.
It turns out that the soliton with
the velocity $c_-$ always propagates to the left whereas the soliton with the velocity $c_+$ propagates to the right and left depending on the 
value of $\rho_0 k$. Thus,  the  two-soliton solution exhibits both the overtaking and head-on collisions. \par
Using (2.24), the expressions $(2.23c)$ and $(2.23d)$ become
$${\rm e}^{-\phi_j}={|(1-\rho_0k_j)c_j-\rho_0\kappa^2|\over \rho_0\sqrt{\kappa^4+\rho_0^2}}
={\{(1-\rho_0k_j)c_j-\rho_0\kappa^2\}{\rm sgn}\ c_j,\over \rho_0\sqrt{\kappa^4+\rho_0^2}}, \eqno(2.28)$$
$${\rm e}^{-{\rm i}\psi_j}={\kappa^2c_j+\rho_0^3-{\rm i}\rho_0^2k_jc_j\over \sqrt{\kappa^4+\rho_0^2}\,|c_j|}. \eqno(2.29)$$
where the last expression in (2.28) is obtained by employing (2.24) again with ${\rm sgn}$ being the sign function.
Substituting $c_j$ from (2.24) into (2.28), one can show that ${\rm e}^{-\phi_j}<1$ and hence $\phi_j>0$.
In view of  the relation  $d_j^2-d_l^2=\rho_0^4(-k_j^2+k_l^2)$ 
which comes from   $(2.24b)$, one can derive the formula
$$\kappa^2(d_j-d_l)^2+\rho_0^4(k_j\pm k_l)(k_jd_l\pm k_ld_j)+\kappa^2\rho_0^4(k_j\pm k_l)^2$$
$$={1\over 2}(d_j+d_l+2\kappa^2)\left\{(d_j-d_l)^2+\rho_0^4(k_j\pm k_l)^2\right\}. $$
Inserting this into $(2.23b)$, we obtain a simplified expression for it
 $${\rm e}^{\gamma_{jl}}={(d_j-d_l)^2+\rho_0^4(k_j-k_l)^2\over (d_j-d_l)^2+\rho_0^4(k_j+k_l)^2}.  \eqno(2.30)$$
It will be used in proving the $N$-soliton solution. See appendix B. 
\par
\medskip
\noindent {\bf Remark 2.3.} Equation $(1.1a)$ with a term  $-\rho\rho_x$ instead of  $+\rho\rho_x$ coupled with equation $(1.1b)$, i.e. 
$$m_t+um_x+2mu_x-\rho\rho_x=0, \quad \rho_t+(\rho u)_x=0, \eqno(2.31)$$
has been introduced in  purely mathematical contexts [4-6].
 It exhibits peculiar features when compared with features of the system (1.1). In particular, it admits peakons and kinks  as well as smooth solitons [6].
 The smooth $N$-soliton solutions with $N\leq 4$ have been obtained by using the Darboux transformation [28].
   The exact method of solution developed here enables us to construct the general $N$-soliton solution in a simple manner, which we shall summarize  shortly.
The expressions corresponding to  (2.1)-(2.6) follow  by the replacement of the variables in accordance with  the rule $\rho\rightarrow {\rm i}\,\rho\,(\rho_0\rightarrow {\rm i}\,\rho_0), 
 y\rightarrow {\rm i}y, Y\rightarrow {\rm i}Y$
while other variables remain unchanged. The parametric representation of the solutions then takes the form
$$u(y, \tau)=\left({\rm ln}\,{\tilde f\over f}\right)_\tau, \quad
\rho(y, \tau)=\rho_0-\left({\rm ln}\,{\tilde g\over g}\right)_\tau, \quad
x(y, \tau)={y\over \rho_0}+{\rm ln}\,{\tilde f\over f}+d. \eqno(2.32)$$
The tau-functions associated with the $N$-soliton solution can be obtained from (2.21)-(2.23) if one replaces the parameters as 
$k_j\rightarrow -{\rm i}k_j, c_j\rightarrow {\rm i}c_j, y_{j0}\rightarrow {\rm i}y_{j0}\ (j=1, 2, ...., N)$, in addition to the
replacements of the variables prescribed above. The soliton solutions have a rich mathematical structure and their properties 
 deserve further study.  The results of the detailed analysis
will be reported elsewhere.
\par
\bigskip
\noindent {\bf 3. Properties of soliton solutions} \par
\bigskip
\noindent In this section, we first explore the properties of the one-soliton solution in detail and then perform an asymptotic analysis of 
 the general $N$-soliton solution. Consequently, the formula for the phase shift of each soliton will be derived. 
 The two-soliton case is discussed in some detail.\par
 \medskip
\noindent{\it 3.1. One-soliton solution} \par
\medskip
\noindent The tau-functions corresponding to the one-soliton solution are given by (2.21)  and (2.22) with $N=1$
$$f=1+{\rm e}^{\xi+\phi}, \qquad \tilde f=1+{\rm e}^{\xi-\phi}, \eqno(3.1)$$
$$g=1+{\rm e}^{\xi+{\rm i}\psi}, \qquad \tilde g=1+{\rm e}^{\xi-{\rm i}\psi}, \eqno(3.2)$$
with 
$$\xi=k\left(y-c\tau -y_0\right), \eqno(3.3a)$$
$$c=c_{\pm}={\rho_0^3\over \pm\sqrt{\kappa^4+\rho_0^2-\rho_0^4k^2}- \kappa^2}, \eqno(3.3b)$$
$${\rm e}^{-\phi}={|(1-\rho_0k)c-\rho_0\kappa^2|\over \rho_0\sqrt{\kappa^4+\rho_0^2}}, \eqno(3.3c)$$
$${\rm e}^{-{\rm i}\psi}={\kappa^2c+\rho_0^3-{\rm i}\rho_0^2kc\over \sqrt{\kappa^4+\rho_0^2}\,|c|}, \eqno(3.3d)$$
where we have put $\xi=\xi_1, k=k_1, c=c_1, \phi=\phi_1, \psi=\psi_1$ and $y_0=y_{10}$ for simplicity. \par
The parametric representation of the one-soliton solution is obtained by introducing (3.1) and (3.2) with (3.3) into (2.16)-(2.18). 
It can be written in the form
$$u={kc\, \sinh\,\phi\over \cosh\,\xi+\cosh\,\phi}, \eqno(3.4a)$$
$$\rho=\rho_0+{kc\,\sin\,\psi\over \cosh\,\xi+\cos\,\psi}, \eqno(3.4b)$$
$$X\equiv x-\tilde c t-x_0={\xi\over \rho_0k}+{\rm ln}{1-\tanh\,{\phi\over 2}\,\tanh{\xi\over 2}\over 1+\tanh\,{\phi\over 2}\,\tanh{\xi\over 2}}, \eqno(3.4c)$$
with
$$\sinh\,\phi={k|c|\over \sqrt{\kappa^4+\rho_0^2}},\qquad \cosh\,\phi=\sqrt{1+{k^2c^2\over \kappa^4+\rho_0^2}}, \eqno(3.4d)$$
$$\sin\,\psi= {\rho_0^2kc\over \sqrt{\kappa^4+\rho_0^2}\,|c|}, \qquad \cos\,\psi={\kappa^2c+\rho_0^3\over \sqrt{\kappa^4+\rho_0^2}\,|c|}, \eqno(3.4e)$$
where $\tilde c=c/\rho_0$ is the velocity of the soliton in the $(x, t)$ coordinate system, $x_0=y_0/\kappa$ and the constant $d$ in (2.18) has been chosen
such that $\xi=0$ corresponds to $X=0$.
The traveling wave coordinate $X$ defined by (3.4c) is particularly useful for the description of the one-soliton solution since it becomes stationary in this
coordinate system. One can use the formula $\tanh(\phi/2)=\sinh\,\phi/(\cosh\,\phi+1)$ to rewrite (3.4c) in terms of $\sinh\,\phi$ and $\cosh\,\phi$. 
\par
It now follows from $(3.4d)$ and $(3.4e)$ that $c\,\sin\,\psi=\rho_0^2\,\sinh\,\phi$. Since $\phi>0$, the sign of $c$ must coincide with that of $\sin\,\psi$.
This condition coupled with $(3.4e)$ is used to determine the permissible  value of $\psi$. Explicitly,
$$c_+\, (0<\rho_0k<1): 0<\psi<\pi/2, \quad c_+\, (1<\rho_0k<\sqrt{\kappa^4+\rho_0^2}/\rho_0): \pi<\psi<3\pi/2,$$ 
$$c_-\,(0<\rho_0k<\sqrt{\kappa^4+\rho_0^2}/\rho_0): 3\pi/2<\psi<2\pi. \eqno(3.4f)$$
\par
Let us now describe some important properties of the solution. \par
\medskip
\noindent {\it (a) Smoothness of the solution} \par
\medskip
\noindent We compute the $y$ derivative of $x$ from (3.4c) to obtain
$$x_y={1\over \rho_0}-{k\,\sinh\,\phi\over \cosh\,\xi+\cosh\,\phi}. \eqno(3.5)$$
Since $k>0$ and $\phi>0$, one has the inequality $x_y\geq x_y|_{\xi=0}$. 
Substituting  $(3.4d)$ for $\sinh\,\phi$ and $\cosh\,\phi$  and using $(2.23e)$, we obtain
$$x_y|_{\xi=0}={1\over \rho_0}-{k\,\sinh\,\phi\over 1+\cosh\,\phi}$$
$$={1\over \rho_0}\left[1-{1\over |c|}\left(|c-\rho_0\kappa^2|-\rho_0\sqrt{\kappa^4+\rho_0^2}\right)\right]
={1\over |c|}\left(\sqrt{\rho_0^2+\kappa^4}+\kappa^2{\rm sgn}\,c\right). \eqno(3.6)$$
  The last  expression follows from the previous one by considering
the cases $c>0$ and $c<0$ separately with the help of  the inequalities (2.26) and (2.27) for $c_\pm$. Note, in particular that $c_+>\rho_0\kappa^2$ for $0<\rho_0k<1$
which is a unique positive branch of the dispersion curve, as is evident from Figure 1. 
Thus, if $c$ is finite, then $x_y>0$, and the map (2.1) becomes one-to-one, assuring that the solution is smooth and nonsingular. 
Actually, one can show that the derivatives $du/dX$ and $d\rho /dX$ are finite for arbitrary $X \in \mathbb{R}$. 
Furthermore, it turns out from $(3.3b)$ and (3.6) that  the smoothness of the solution prevails in the zero dispersion limit $\kappa \rightarrow 0$.  However, 
the limit operation $\rho_0 \rightarrow 0$ with $\kappa$ being fixed at a constant value requires  a delicate analysis. See section 4.1.
\par
\medskip
\noindent {\it (b) Amplitude-velocity relation} \par
\medskip
\noindent The amplitude-velocity relation of the soliton  is an important characteristic of the wave.  
It can be derived simply from the explicit form  (3.4) of the solution. To this end,
let $A_\rho$  be the amplitude of the wave  measured from the constant level $\rho=\rho_0$ and $A_u$ be that of the fluid velocity, i.e.,
$A_\rho=\rho(X=0)-\rho_0$, and $ A_u=|u(X=0)|$.  We find that
$$A_\rho=\left(\sqrt{\kappa^4+\rho_0^2}\,|\tilde c|-\kappa^2\tilde c-\rho_0^2\right)/\rho_0,  \eqno(3.7a)$$
$$A_u=\left ||\tilde c-\kappa^2|-\sqrt{\kappa^4+\rho_0^2}\right |, \eqno(3.7b)$$
where $\tilde c=c/\rho_0$.  Note  that
$$u(X=0)=kc\,\tanh\, {\phi\over 2}=\left(|\tilde c-\kappa^2|-\sqrt{\kappa^4+\rho_0^2}\right)\, {\rm sgn}\,\tilde c. $$
Invoking  the expression of the velocity $c$ from  $(3.3b)$, we can see that $A_\rho>0$  for arbitrary $c=c_\pm$ whereas $u(X=0)>0$
for $c>0$ and  $u(X=0)<0$ for $c<0$.
These results show that the profile of $\rho$ is always of bright type, but that of $u$ depends on the propagation direction
of the soliton. Actually, if $c$ is positive (negative), then $u$ is curved upward (downward). 
 \par
Figure 2 depicts the typical profile of $u$ and $\rho$  for  the right-going soliton (a), and the left-going soliton (b) and (c), respectively \par
\bigskip

\begin{figure}[t]
\begin{center}
\includegraphics[width=16cm]{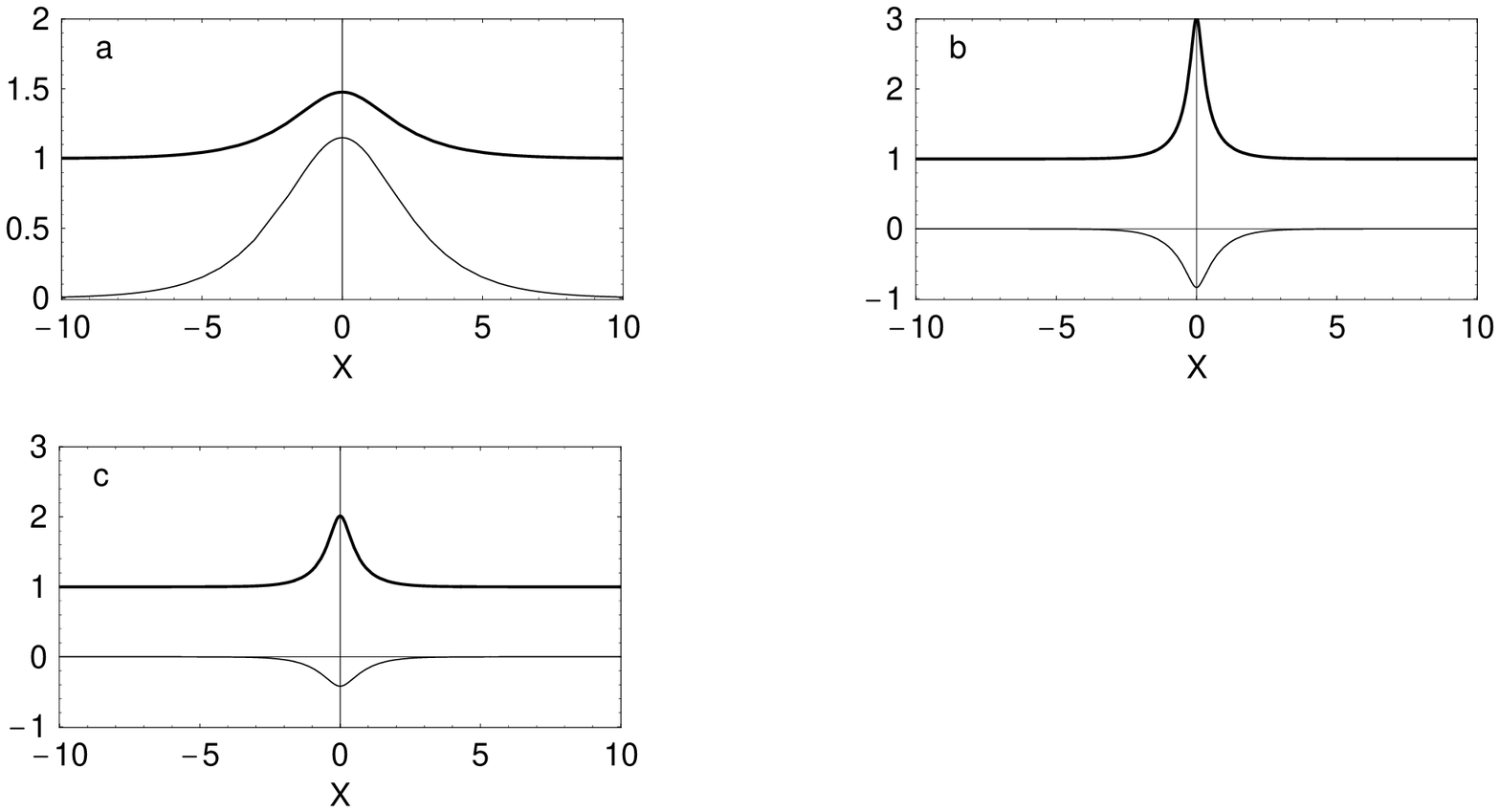}
\end{center}
{{\bf Figure 2.}\ One-soliton solution. $u$: thin solid curve, \ $\rho$: bold solid curve. a: $\kappa=1, \rho_0=1, k=0.4, \tilde c=\tilde c_+=2.81 $, 
\ b: $\kappa=1, \rho_0=1, k=1.4, \tilde c=\tilde c_+=-1.25  $, c: $\kappa=1, \rho_0=1, k=1.4, \tilde c=\tilde c_-= -0.83 .$}
\end{figure}

\noindent{\it 3.2. $N$-soliton solution} \par
\bigskip
\noindent Here, we investigate the asymptotic behavior of the $N$-soliton solution  for large time. Let $\tilde c_n(=c_n/\rho_0),\ (n=1, 2, ..., N)$ be the 
velocity of the $n$th soliton in the $(x, t)$ coordinate system, and order them in accordance with the relation
$\tilde c_N<\tilde c_{N-1}<... < \tilde c_1$.
We take the limit $t\rightarrow -\infty$  with the phase variable $\xi_n$ of the $n$th soliton being fixed.
Then, the other  phase variables behave like $\xi_1, \xi_2, ..., \xi_{n-1} \rightarrow +\infty$, and $\xi_{n+1}, \xi_{n+2}, ..., \xi_N \rightarrow -\infty$.
Performing an asymptotic analysis for the tau-functions (2.21) and (2.22),
the leading-order approximations for them  are found to be
$$ f \sim \left(\prod_{1\leq j<l\leq n-1}{\rm e}^{\gamma_{jl}}\right)\, {\rm exp}\left[\sum_{j=1}^{n-1}(\xi_j+\phi_j)\right]\left(1+{\rm e}^{\xi_n+\phi_n+\delta_n^{(-)}}\right), \eqno(3.8a)$$
$$\tilde f \sim \left(\prod_{1\leq j<l\leq n-1}{\rm e}^{\gamma_{jl}}\right)\, {\rm exp}\left[\sum_{j=1}^{n-1}(\xi_j-\phi_j)\right]\left(1+{\rm e}^{\xi_n-\phi_n+\delta_n^{(-)}}\right), \eqno(3.8b)$$
$$ g \sim \left(\prod_{1\leq j<l\leq n-1}{\rm e}^{\gamma_{jl}}\right)\, {\rm exp}\left[\sum_{j=1}^{n-1}(\xi_j+{\rm i}\psi_j)\right]\left(1+{\rm e}^{\xi_n+{\rm i}\psi_n+\delta_n^{(-)}}\right), \eqno(3.9a)$$
$$ \tilde g \sim \left(\prod_{1\leq j<l\leq n-1}{\rm e}^{\gamma_{jl}}\right)\, {\rm exp}\left[\sum_{j=1}^{n-1}(\xi_j-{\rm i}\psi_j)\right]\left(1+{\rm e}^{\xi_n-{\rm i}\psi_n+\delta_n^{(-)}}\right), \eqno(3.9b)$$
where
$$\delta_n^{(-)}=\sum_{j=1}^{n-1}\gamma_{nj}=\sum_{j=1}^{n-1}{\rm ln}\left[{(d_n-d_j)^2+\rho_0^4(k_n-k_j)^2\over (d_n-d_j)^2+\rho_0^4(k_n+k_j)^2}\right]. \eqno(3.10)$$
Substituting (3.8) and (3.9) into (2.16)-(2.18), we obtain the asymptotic form of $u$, $\rho$ and $x$
$$u \sim {k_nc_n\, \sinh\,\phi_n\over \cosh\left(\xi_n+\delta_n^{(-)}\right)+\cosh\,\phi_n}, \eqno(3.11)$$
$$\rho \sim \rho_0+{k_nc_n\,\sin\,\psi_n\over \cosh\left(\xi_n+\delta_n^{(-)}\right)+\cos\,\psi_n}, \eqno(3.12)$$
$$x-\tilde c_nt-x_{n0} \sim {\xi_n\over \rho_0k_n} +{\rm ln}\ {1-\tanh\,{\phi_n\over 2}\,\tanh{\left(\xi_n+\delta_n^{(-)}\right)\over 2}\over 1+\tanh\,{\phi_n\over 2}\,\tanh{\left(\xi_n+\delta_n^{(-)}\right)\over 2}}
-2\sum_{j=1}^{n-1}\phi_j. \eqno(3.13)$$
\par
In the limit $t \rightarrow +\infty$, on the other hand, we see that $\xi_1, \xi_2, ..., \xi_{n-1} \rightarrow -\infty$, and $\xi_{n+1}, \xi_{n+2}, ..., \xi_N \rightarrow +\infty$.
Applying the similar analysis  yields the asymptotic forms
corresponding to (3.8)-(3.13)
$$ f \sim \left(\prod_{n+1\leq j<l\leq N}{\rm e}^{\gamma_{jl}}\right)\, {\rm exp}\left[\sum_{j=n+1}^{N}(\xi_j+\phi_j)\right]\left(1+{\rm e}^{\xi_n+\phi_n+\delta_n^{(+)}}\right), \eqno(3.14a)$$
$$\tilde f \sim \left(\prod_{n+1\leq j<l\leq N}{\rm e}^{\gamma_{jl}}\right)\, {\rm exp}\left[\sum_{j=n+1}^{N}(\xi_j-\phi_j)\right]\left(1+{\rm e}^{\xi_n-\phi_n+\delta_n^{(+)}}\right), \eqno(3.14b)$$
$$ g \sim \left(\prod_{n+1\leq j<l\leq N}{\rm e}^{\gamma_{jl}}\right)\, {\rm exp}\left[\sum_{j=n+1}^{N}(\xi_j+{\rm i}\psi_j)\right]\left(1+{\rm e}^{\xi_n+{\rm i}\psi_n+\delta_n^{(+)}}\right), \eqno(3.15a)$$
$$ \tilde g \sim \left(\prod_{n+1\leq j<l\leq N}{\rm e}^{\gamma_{jl}}\right)\, {\rm exp}\left[\sum_{j=n+1}^{N}(\xi_j-{\rm i}\psi_j)\right]\left(1+{\rm e}^{\xi_n-{\rm i}\psi_n+\delta_n^{(+)}}\right), \eqno(3.15b)$$
where
$$\delta_n^{(+)}=\sum_{j=n+1}^{N}\gamma_{nj}=\sum_{j=n+1}^{N}{\rm ln}\left[{(d_n-d_j)^2+\rho_0^4(k_n-k_j)^2\over (d_n-d_j)^2+\rho_0^4(k_n+k_j)^2}\right], \eqno(3.16)$$
and
$$u \sim {k_nc_n\, \sinh\,\phi_n\over \cosh\left(\xi_n+\delta_n^{(+)}\right)+\cosh\,\phi_n}, \eqno(3.17)$$
$$\rho \sim \rho_0+{k_nc_n\,\sin\,\psi_n\over \cosh\left(\xi_n+\delta_n^{(+)}\right)+\cos\,\psi_n}, \eqno(3.18)$$
$$x-\tilde c_nt-x_{n0} \sim {\xi_n\over \rho_0k_n} +{\rm ln}\ {1-\tanh\,{\phi_n\over 2}\,\tanh{\left(\xi_n+\delta_n^{(+)}\right)\over 2}\over 1+\tanh\,{\phi_n\over 2}\,\tanh{\left(\xi_n+\delta_n^{(+)}\right)\over 2}}
-2\sum_{j=n+1}^N\phi_j. \eqno(3.19)$$
\par
These results show that as $t \rightarrow \pm\infty$, the $N$-soliton solution is represented by a superposition of $N$ independent solitons each of which has  the form 
of the one-soliton solution given by (3.4).
The net effect of the collision of solitons appears as a phase shift. To see this, let $x_{nc}$ be the center position of the $n$th soliton. It then follows from (3.13) and (3.19) that the trajectory of $x_{nc}$ is
given by
$$x_{nc} \sim \tilde c_nt-{\delta_n^{(-)}\over \rho_0k_n}-2\sum_{j=1}^{n-1}\phi_j, \quad (t \rightarrow -\infty), \eqno(3.20a)$$
$$x_{nc} \sim \tilde c_nt-{\delta_n^{(+)}\over \rho_0k_n}-2\sum_{j=n+1}^{N}\phi_j, \quad (t \rightarrow +\infty). \eqno(3.20b)$$
We define the phase shift of the $n$th soliton which propagates to the right by $\Delta_n^{R}=x_{nc}(t\rightarrow +\infty)-x_{nc}(t\rightarrow -\infty)$, and that propagates to the left by
$\Delta_n^{L}=x_{nc}(t\rightarrow -\infty)-x_{nc}(t\rightarrow +\infty)$. Using (2.23c), (3.10), (3.16) and (3.20), we find that
$$\Delta_n^{R}={1\over \rho_0k_n} \left[\sum_{j=1}^{n-1}{\rm ln}\left[{(d_n-d_j)^2+\rho_0^4(k_n-k_j)^2\over (d_n-d_j)^2+\rho_0^4(k_n+k_j)^2}\right]
-\sum_{j=n+1}^{N}{\rm ln}\left[{(d_n-d_j)^2+\rho_0^4(k_n-k_j)^2\over (d_n-d_j)^2+\rho_0^4(k_n+k_j)^2}\right]\right]$$
$$+\sum_{j=n+1}^N{\rm ln}\left[(1-\rho_0k_j)\tilde c_j-\kappa^2\over (1+\rho_0k_j)\tilde c_j-\kappa^2\right]
-\sum_{j=1}^{n-1}{\rm ln}\left[(1-\rho_0k_j)\tilde c_j-\kappa^2\over (1+\rho_0k_j)\tilde c_j-\kappa^2\right]. \eqno(3.21)$$
The expression of $\Delta_n^{L}$ is equal to $-\Delta_n^{R}$. \par
\bigskip
\noindent{\it 3.3. Two-soliton solution} \par
\bigskip
\noindent The two-soliton solution is the most fundamental element in understanding the dynamics of solitons since each soliton exhibits pair wise interactions with every other soliton, as
indicated by the formulas of the phase shift.
There exist two types of interactions for the CH2 system, i.e., the overtaking  and head-on collisions.  We describe them separately. \par
The tau-functions for the two-soliton solution are given by (2.21)-(2.23) and (2.30)  with $N=2$. They read
$$f=1+{\rm e}^{\xi_1+\phi_1}+{\rm e}^{\xi_2+\phi_2}+\delta\, {\rm e}^{\xi_1+\xi_2+\phi_1+\phi_2}, \eqno (3.22a)$$
$$\tilde f=1+{\rm e}^{\xi_1-\phi_1}+{\rm e}^{\xi_2-\phi_2}+\delta\, {\rm e}^{\xi_1+\xi_2-\phi_1-\phi_2}, \eqno (3.22b)$$
$$g=1+{\rm e}^{\xi_1+{\rm i}\psi_1}+{\rm e}^{\xi_2+{\rm i}\psi_2}+\delta\, {\rm e}^{\xi_1+\xi_2+{\rm i}\psi_1+{\rm i}\psi_2}, \eqno (3.23a)$$
$$\tilde g=1+{\rm e}^{\xi_1-{\rm i}\psi_1}+{\rm e}^{\xi_2-{\rm i}\psi_2}+\delta\, {\rm e}^{\xi_1+\xi_2-{\rm i}\psi_1-{\rm i}\psi_2}, \eqno (3.23b)$$

\begin{figure}[h]
\begin{center}
\includegraphics[width=16cm]{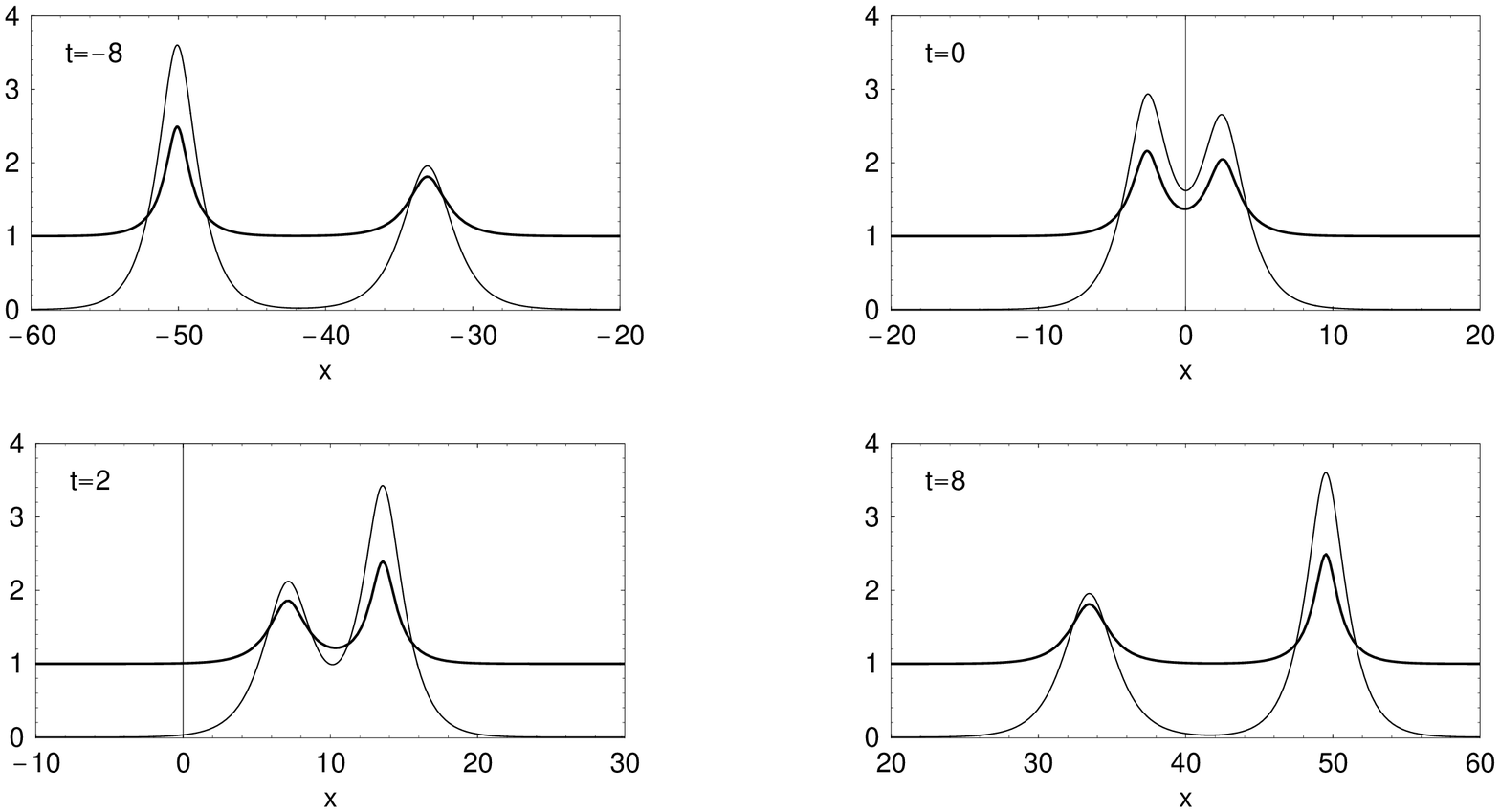}
\end{center}
{{\bf Figure 3.}\ The overtaking collision of two solitons. $u$: thin solid curve, \ $\rho$: bold solid curve. $\kappa=1, \rho_0=1, k_1=0.8, k_2=0.7, \tilde c_{1+}=6.02, \tilde c_{2+}=4.37$.}
\end{figure}
where
$$\xi_j=k_j\left(y-c_j\tau-y_{j0}\right), \qquad (j=1, 2),\eqno(3.24a)$$
$$\delta={\rm  e}^{\gamma_{12}}={(d_1-d_2)^2+\rho_0^4(k_1-k_2)^2\over (d_1-d_2)^2+\rho_0^4(k_1+k_2)^2}, \eqno(3.24b)$$
$${\rm e}^{-\phi_j}=\sqrt{(1-\rho_0k_j)c_j-\rho_0\kappa^2\over (1+\rho_0k_j)c_j-\rho_0\kappa^2},\qquad (j=1, 2),\eqno(3.24c)$$
$${\rm e}^{-{\rm i}\psi_j}=\sqrt{\left({\kappa^2\over \rho_0}-{\rm i}\rho_0k_j\right)c_j+\rho_0^2\over \left({\kappa^2\over \rho_0}+{\rm i}\rho_0k_j\right)c_j+\rho_0^2},\qquad (j=1, 2).\eqno(3.24d)$$
Recall from (2.24) that the velocity of $j$th soliton in $(x, t)$ coordinate system is given by
$$\tilde c_j=c_j/\rho_0={\rho_0^2\over d_j-\kappa^2}, \quad  d_j=\epsilon_j\sqrt{\kappa^4+\rho_0^2-\rho_0^4k_j^2}, \quad (j=1, 2). \eqno(3.25)$$
Substituting (3.22)-(3.25) into (2.16)-(2.18), we obtain the parametric representation of the two-soliton solution.
Since the velocity $\tilde c_j$ takes either the positive or negative values,
 this solution enables us to describe  both the overtaking and head-on collisions  between two solitons. \par
\bigskip

\begin{figure}[h]
\begin{center}
\includegraphics[width=16cm]{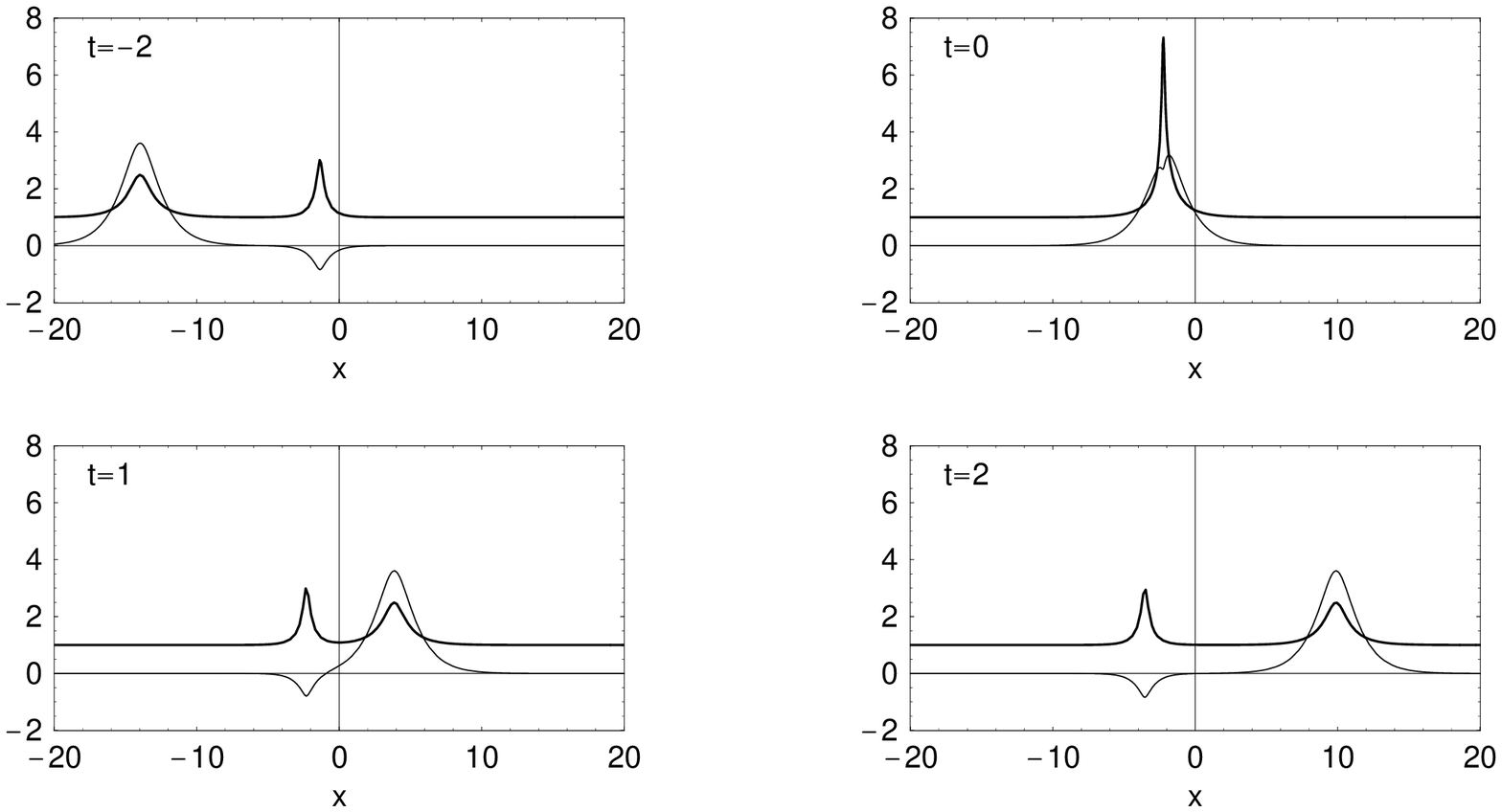}
\end{center}
{{\bf Figure 4.}\ The head-on collision of two solitons. $u$: thin solid curve, \ $\rho$: bold solid curve. $\kappa=1, \rho_0=1, k_1=0.8, k_2=1.4
, \tilde c_{1+}=6.02, \tilde c_{2+}=-1.25$}
\end{figure}

\noindent {\it (a) Overtaking collision} \par
\medskip
\noindent We consider the case $c_j=c_{j+}, 0<\rho_0k_j<1$ so that $0<\tilde c_{2+}<\tilde c_{1+}$. Figure 3 illustrates the overtaking collision of two solitons for four distinct values of $t$.
 The solitonic feature of the solution is obvious from the figure which confirms an asymptotic analysis presented in \S 3.1. The phase shift of each soliton is given
 by (3.21). Explicitly, 
 $$\Delta_1^R=-{1\over \rho_0k_1}\,{\rm ln}\ \left[{(d_1-d_2)^2+\rho_0^4(k_1-k_2)^2\over (d_1-d_2)^2+\rho_0^4(k_1+k_2)^2}\right]
+\,{\rm ln}\ \left[{(1-\rho_0k_2)\tilde c_2-\kappa^2\over (1+\rho_0k_2)\tilde c_2-\kappa^2}\right], \eqno(3.26a)$$
 $$\Delta_2^R={1\over \rho_0k_2}\,{\rm ln}\ \left[{(d_1-d_2)^2+\rho_0^4(k_1-k_2)^2\over (d_1-d_2)^2+\rho_0^4(k_1+k_2)^2}\right]
-{\rm ln}\ \left[{(1-\rho_0k_1)\tilde c_1-\kappa^2\over (1+\rho_0k_1)\tilde c_1-\kappa^2}\right], \eqno(3.26b)$$
with
$$d_1=\sqrt{\kappa^4+\rho_0^2-\rho_0^4k_1^2}, \quad d_2=\sqrt{\kappa^4+\rho_0^2-\rho_0^4k_2^2}. \eqno(3.26c)$$
\medskip
\noindent {\it (b) Head-on collision} \par
\medskip
\noindent An example of the head-on collision is shown in Figure 4, where the velocity of each soliton is chosen as $c_{2+}<0<c_{1+}$. The formula of the phase shift for the right-running soliton 
is the same as $(3.26a)$ whereas that of the left-running soliton is given by $\Delta_2^L=-\Delta_2^R$.  \par

\bigskip
\noindent {\bf Remark 3.1.}\par
\medskip
\noindent As noticed in [7], the CH2 system (1.1) with $\kappa=0$ does not admit peakons. 
The same will be true in the case of $\kappa\not= 0$.
 Recall, however that another integrable CH2 system (2.31) exhibits peakons when the parameter $\kappa$ is related to the boundary value $\rho_0$ of $\rho$
as $\rho_0=\kappa^2$. See, for example [28]. \par
\bigskip
\noindent {\bf 4. Reductions to the CH equation,  the HS2 system and the HS equation} \par
\bigskip
\par
{\large
$$\begin{CD}
\fboxrule=0.3mm\fbox{CH2} @>{\rm SL}>> \fboxrule=0.3mm\fbox{CH} \\
@VV {\rm SWL}V @VV {\rm SWL}V \\
\fboxrule=0.3mm\fbox{HS2} @>{\rm SL}>> \fboxrule=0.3mm\fbox{HS}
\end{CD}$$}
\par
\noindent {\bf Figure 5.} The reduction process for the CH2 system in which  SL and SWL abbraviate \par
\vspace{-2mm}
\noindent the scaling and short-wave limits, respectively. 
\par
\bigskip
\noindent In this section, we first show that the CH2 system and its $N$-soliton solution reduce to  the CH equation 
and the corresponding $N$-soliton solution under an appropriate limiting 
procedure, or more precisely, the scaling limit. Then, we demonstrate that the short-wave limit of the CH2 system yields the HS2 system. 
The reduction to the HS equation is outlined shortly. \par
The primary difference between the scaling limit and short-wave limit is that in the former limit, no scalings are prescribed  for the space and time variables
whereas in the latter limit, the rapidly-varying space variable $\hat x$ and slowly-varying time variable $\hat t$ are
introduced via the relations $\hat x=x/\epsilon$ and $\hat t=\epsilon t$, where $\epsilon$ is a scaling parameter. 
The reduction process developed here is displayed in Figure 5 in which the two different avenues  
 leading to the HS equation are indicated.
\par
\bigskip
\noindent{\it 4.1. Reduction to the CH equation} \par
\bigskip
\noindent The CH equation (1.2) is derived formally from the CH2 system by putting $\rho=0$.  In this setting, one must impose the boundary condition $\rho_0=0$.
The $N$-soliton solution of the CH equation is reduced from that of the CH2 system by taking the limit $\rho_0\rightarrow 0$. This limiting procedure is, however highly non-trivial, as will be
shown below. \par
First, we introduce the following scaling variables with an overbar
$$u=\bar u,\ \rho=\rho_0\bar\rho,\ m=\bar m, \ x=\bar x, \ y={\rho_0\over \kappa}\,\bar y,\ t=\bar t, \  \tau=\bar\tau, \ d=\bar d,$$
$$\ k_j={\kappa\over \rho_0}\,\bar k_j, \ c_j={\rho_0\over \kappa}\,\bar c_j,\ y_{j0}={\rho_0\over \kappa}\,\bar y_{j0}, \ (j=1, 2, ..., N). \eqno(4.1)$$
Then, the leading-order asymptotics of $c_j$ from (2.24), and  $\gamma_{jl}, \phi_j$, and $\psi_j$  from (2.23) are found to be
$$c_j \sim {2\rho_0\kappa^2\over 1-(\kappa \bar k_j)^2}, \ (j=1,2, ..., N), \eqno(4.2a)$$
$${\rm e}^{\gamma_{jl}}=\left({\bar k_j-\bar k_l\over \bar k_j+\bar k_l}\right)^2\equiv {\rm e}^{\bar \gamma_{jl}}, \ (j, l=1, 2, ..., N; j\not=l), \eqno(4.2b)$$
$${\rm e}^{-\phi_j} \sim {1-\kappa \bar k_j\over 1+\kappa \bar k_j}\equiv {\rm e}^{-\bar\phi_j}, \ (j=1,2, ..., N), \eqno(4.2c)$$
$${\rm e}^{-{\rm i}\psi_j} \sim 1-{\rm i}\, {\rho_0\over \kappa}\, \bar k_j,  \ (j=1,2, ..., N). \eqno(4.2d)$$
We note that a limiting form $\bar c_j \sim -\rho_0^2/(2\kappa)$ of the velocity which arises from (2.24) with $\epsilon_j=-1\ (j=1, 2)$ is not relevant since 
in accodance with the scaling (4.1), this expression leads to $\bar c_j/\rho_0 \sim -\rho_0/(2\kappa) \rightarrow 0\ (\rho_0 \rightarrow 0)$, showing that
the velocity in the $(\bar x, \bar t)$ coordinate system degenerates to zero.
\par
The  asymptotics  of the tau-functions  $f$ and $\tilde f$ from (2.21) and
$g$ and $\tilde g$ from (2.22) become
$$f \sim \sum_{\mu=0,1}{\rm exp}\left[\sum_{j=1}^N\mu_j\left(\bar\xi_j+\bar\phi_j\right)
+\sum_{1\le j<l\le N}\mu_j\mu_l\bar\gamma_{jl}\right] \equiv \bar f, \eqno(4.3a)$$
$$\tilde f \sim \sum_{\mu=0,1}{\rm exp}\left[\sum_{j=1}^N\mu_j\left(\bar\xi_j-\bar\phi_j\right)
+\sum_{1\le j<l\le N}\mu_j\mu_l\bar\gamma_{jl}\right] \equiv \bar{\tilde f}, \eqno(4.3b)$$
$$g =\bar f_0+{\rm i}\,{\rho_0\over \kappa}\,\bar f_{0,\bar y}+O(\rho_0^2), \eqno(4.4a)$$
$$\tilde g =\bar f_0-{\rm i}\,{\rho_0\over \kappa}\,\bar f_{0,\bar y}+O(\rho_0^2), \eqno(4.4b)$$
where
$$\bar f_0 = \sum_{\mu=0,1}{\rm exp}\left[\sum_{j=1}^N\mu_j\bar\xi_j
+\sum_{1\le j<l\le N}\mu_j\mu_l\bar\gamma_{jl}\right], \eqno(4.5a)$$
$$\bar\xi_j=\bar k_j\left(\bar y-\bar c_j\bar\tau-\bar y_{j0}\right), \quad \bar c_j={2\kappa^3\over 1-(\kappa \bar k_j)^2},  \qquad (j=1, 2, ..., N).\eqno(4.5b)$$
Introducing (4.1), (4.3) and (4.4) into (2.16)-(2.18) and taking the limit $\rho_0\rightarrow 0$, we obtain the limiting forms of $u$, $\rho$ and $x$
$$\bar u=\left({\rm ln}\,{\bar{\tilde f}\over \bar f}\right)_{\bar\tau}, \eqno(4.6)$$
$$\rho \sim \rho_0\left(1-{2\over\kappa}\,({\rm ln} \bar f_0)_{\bar y\tau}\right)\equiv \rho_0\bar\rho, \eqno(4.7)$$
$$\bar x={\bar y\over\kappa}+ {\rm ln}\,{\bar{\tilde f}\over \bar f}+\bar d. \eqno(4.8)$$
\par
The parametric representation of the $N$-soliton solution given by (4.6) and (4.8) with the tau-functions (4.3)
coincides perfectly with that  of the CH equation presented in [21].  In particular,  the one-soliton solution (3.4) reduces to
$$\bar u={2\kappa\bar c\bar k^2\over 1+\kappa^2\bar k^2+(1-\kappa^2\bar k^2)\,\cosh\,\bar\xi}, \eqno(4.9a)$$
$$\bar X=\bar x-\bar{\tilde c}-\bar x_0={\bar \xi\over \kappa\bar k}
+{\rm ln}\,{(1-\kappa\bar k)\,{\rm e}^{\bar \xi}+1+\kappa\bar k \over (1+\kappa\bar k)\,{\rm e}^{\bar \xi}+1-\kappa\bar k}, \eqno(4.9b)$$
with
$$\bar\xi=\bar k(\bar y-\bar c\bar\tau-\bar y_0) \quad \bar c={2\kappa^3\over 1-(\kappa\bar k)^2}, \quad \bar{\tilde c}=\bar c/\kappa, \eqno(4.9c)$$
reproducing the one-soliton solution of the CH equation. \par
The limiting form of the phase shift  which is denoted by $\bar\Delta_n^R$ can be derived from  (3.21) by using $(4.2a)$. It reads
$$\bar\Delta_n^R={1\over \kappa\bar k_n}\left[\sum_{j=1}^{n-1}{\rm ln}\left({\bar k_n-\bar k_j\over \bar k_n+\bar k_j}\right)^2
-\sum_{j=n+1}^N{\rm ln}\left({\bar k_n-\bar k_j\over \bar k_n+\bar k_j}\right)^2\right]$$
$$+\sum_{j=n+1}^N{\rm ln}\left({1-\kappa \bar k_j\over 1+\kappa \bar k_j}\right)^2-\sum_{j=1}^{n-1}{\rm ln}\left({1-\kappa \bar k_j\over 1+\kappa \bar k_j}\right)^2. \eqno(4.10)$$
This is just the formula for the phase shift of the $N$-soliton solution of the CH equation presented in [21]. \par
\bigskip
\noindent{\bf Remark 4.1.}\par
\noindent If we put $\bar r=\kappa-2({\rm ln}\,\bar f_0)_{\bar y\tau}$, then 
$$\bar\rho={\bar r\over \kappa}, \quad \bar m=\bar r^2. \eqno(4.11) $$ 
The first equation in (4.11) follows immediately from (4.7), and the second equation can be derived by taking
the scaling limit of (2.20). 
The reciprocal
transformation $(2.1a)$  reproduces the corresponding one for the CH equation [21]
$$d\bar y=\bar r\,d\bar x-\bar r\bar u\,d\bar t, \quad d\bar \tau=d\bar t. \eqno(4.12)$$
In terms of the scaling variables (4.1), 
the bilinear equations (2.11)-(2.13) reduce respectively to 
$$\kappa D_{\bar y}\bar{\tilde f}\cdot \bar f+\bar{\tilde f}\bar f-\bar f_0^2=0, \eqno(4.13)$$
$$D_{\bar\tau}D_{\bar y}\bar f_0\cdot \bar f_0+\kappa(\bar{\tilde f}\bar f-\bar f_0^2)=0, \eqno(4.14)$$
$$\kappa D_{\bar\tau}D_{\bar y} \bar{\tilde f}\cdot \bar f +D_{\bar\tau}\bar{\tilde f}\cdot \bar f+\kappa^3 D_{\bar y} \bar{\tilde f}\cdot \bar f=0. \eqno(4.15)$$
The scaling limit of (2.14) is performed after eliminating the derivative $D_\tau\tilde g\cdot g$ in (2.14) by means of  (2.12).
We then find that the limiting form of (2.14) coincides with (4.14).
One can show that the tau-functions $\bar f$ and $\bar{\tilde f}$ from (4.3) and $\bar f_0$ from (4.5) solve the above bilinear equations. \par
\bigskip
\noindent{\it 4.2. Reduction to the HS2 system} \par
\bigskip
\noindent The HS2 system arises from the short-wave limit of the CH2 system. In this case, we  introduce the scaling variables with a hat
$$u=\epsilon^2\hat u,\ \rho=\epsilon\hat\rho,\ m=\hat m, \ x=\epsilon\hat x, \ y=\epsilon^2 \hat y,\ t={\hat t \over \epsilon}, \  \tau={\hat\tau\over \epsilon}. \eqno(4.16)$$
Rescaling the CH2 system (1.1) by (4.16) and taking the limit $\epsilon \rightarrow 0$, we obtain the HS2 system
$$\hat m_{\hat t}+\hat u\hat m_{\hat x}+2\hat m \hat u_{\hat x}+\hat\rho\hat\rho_{\hat x}=0, \eqno(4.17a)$$
$$\hat \rho_{\hat t}+(\hat\rho \hat u)_{\hat x}=0, \eqno(4.17b)$$
where $\hat m=-\hat u_{\hat x\hat x}+\kappa^2$, which coincides with (1.4) upon removing the hat attached to the variables.
\par
 The $N$-soliton solution of the HS2 system can be recovered from that of the CH2 system by means of a scaling limit.
The appropriate scaling variables are found to be
$$\ k_j={\hat k_j \over \epsilon^2}, \ c_j=\epsilon^3\hat c_j,\ y_{j0}=\epsilon^2\hat y_{j0}, \ (j=1, 2, ..., N), \ \rho_0=\epsilon\hat \rho_0, \  d=\epsilon \hat d. \eqno(4.18)$$
In the limit $\epsilon \rightarrow 0$, the soliton parameters corresponding to those given by (4.2) have the leading-order asymptotics
$$ c_j\sim -{\epsilon^3\over \hat \rho_0\hat k_j^2}\,(\kappa^2+\hat d_j), \quad \hat d_j=\epsilon_j\sqrt{\kappa^4-\hat\rho_0^4\hat k_j^2}, \quad (j=1, 2, ..., N), \eqno(4.19a)$$
$${\rm e}^{\gamma_{jl}} \sim {(\hat d_j-\hat d_l)^2+\hat \rho_0^4(\hat k_j-\hat k_l)^2\over (\hat d_j-\hat d_l)^2+\hat \rho_0^4(\hat k_j+\hat k_l)^2}
\equiv {\rm e}^{\hat \gamma_{jl}}, \ (j, l=1, 2, ..., N; j\not=l), \eqno(4.19b)$$
$${\rm e}^{-\phi_j} \sim  1+\epsilon\,{\hat k_j\hat c_j\over \kappa^2}, \ (j=1,2, ..., N), \eqno(4.19c)$$
$${\rm e}^{-{\rm i}\psi_j} \sim \sqrt{{\left({\kappa^2\over \hat\rho_0}-{\rm i}\hat\rho_0\hat k_j\right)\hat c_j+\hat\rho_0^2 \over
\left({\kappa^2\over \hat \rho_0}+{\rm i}\hat\rho_0\hat k_j\right)\hat c_j+\hat\rho_0^2}} \equiv {\rm e}^{-{\rm i}\hat\psi_j},
 \ (j=1,2, ..., N). \eqno(4.19d)$$

\par
The tau-functions (2.21) and (2.22) have the leading-order asymptotics
$$f \sim \hat f+{\epsilon\over \kappa^2}\,\hat f_{\hat\tau}, \quad \tilde f \sim \hat f-{\epsilon\over \kappa^2}\,\hat f_{\hat\tau}, \eqno(4.20)$$
  $$g \sim \sum_{\mu=0,1}{\rm exp}\left[\sum_{j=1}^N\mu_j\left(\hat\xi_j+{\rm i}\hat\psi_j\right)
+\sum_{1\le j<l\le N}\mu_j\mu_l\hat\gamma_{jl}\right] \equiv \hat g, \eqno(4.21a)$$
 $$\tilde g \sim \sum_{\mu=0,1}{\rm exp}\left[\sum_{j=1}^N\mu_j\left(\hat\xi_j-{\rm i}\hat\psi_j\right)
+\sum_{1\le j<l\le N}\mu_j\mu_l\hat \gamma_{jl}\right] \equiv \hat{\tilde g}, \eqno(4.21b)$$
where
$$\hat f=\sum_{\mu=0,1}{\rm exp}\left[\sum_{j=1}^N\mu_j\hat\xi_j
+\sum_{1\le j<l\le N}\mu_j\mu_l\hat\gamma_{jl}\right], \eqno(4.22a)$$
$$\hat\xi_j=\hat k_j(\hat y-\hat c_j\hat\tau-\hat y_{j0}), 
\quad  \hat c_j= -{1\over \hat \rho_0\hat k_j^2}\,\left(\kappa^2+\epsilon_j\sqrt{\kappa^4-\hat\rho_0^4\hat k_j^2}\right), \quad (j=1, 2, ..., N). \eqno(4.22b)$$  
\par
The parametric representation for the $N$-soliton solution of the HS2 system follows by introducing (4.20) and (4.21) into (2.16)-(2.18) and taking the limit $\epsilon\rightarrow 0$.
Explicitly,
$$\hat u=-{2\over\kappa^2}\,({\rm ln}\, \hat f)_{\hat\tau\hat\tau}, \eqno(4.23)$$
$$\hat\rho=\hat \rho_0-{2\over\kappa^2}\,{\rm i}\left({\rm ln}\,{\hat{\tilde g}\over \hat g}\right)_{\hat\tau}, \eqno(4.24)$$
$$\hat x ={\hat y\over \hat\rho_0}-{2\over\kappa^2}\,({\rm ln}\, \hat f)_{\hat\tau}+\hat d. \eqno(4.25)$$
The limiting forms of (2.19) and (2.20) turn out to be
$${1\over \hat\rho}={1\over \hat\rho_0}-{2\over\kappa^2}\,({\rm ln}\, \hat f)_{\hat\tau\hat y}, \eqno(4.26)$$ 
$${\hat m\over \hat\rho^2}={\kappa^2\over \hat\rho_0^2}+{\rm i}\left({\rm ln}\,{\hat{\tilde g}\over \hat g}\right)_{\hat y}. \eqno(4.27)$$
\par
We write the one-soliton solution for reference.
$$\hat u=-{1\over 2\kappa^2}{(\hat k\hat c)^2\over \cosh^2{\hat\xi\over 2}},
\quad \hat\rho={1\over {1\over\hat\rho_0}+{\hat k^2\hat c\over 2\kappa^2}{1\over \cosh^2{\hat\xi\over 2}}}, \eqno(4.28a)$$
$$\hat X=\hat x-\hat{\tilde c}\hat t-\hat x_0={\hat \xi\over \hat\rho_0\hat k}+{\hat k\hat c\over\kappa^2}\, \tanh\,{\hat\xi\over 2}, \quad \eqno(4.28b)$$
with
$$\hat\xi=\hat k(\hat y-\hat c\hat\tau-\hat y_0), \quad \hat c=-{1\over \hat\rho_0\hat k^2}\left(\kappa^2\pm\sqrt{\kappa^4-\hat\rho_0^4\hat k^2}\right), \quad \hat{\tilde c}=\hat c/\hat\rho_0. \eqno(4.28c)$$
Notice that the velocities $\hat{\tilde c}$ from $(4.28c)$  are negative for both plus and minus signs so that the soliton propagates to the left as opposed to the soliton solution of the CH2 system
for which the bi-directional propagation is possible.  Furthermore, in contrast to the CH2 case, the profile of $\hat\rho$ takes the form of a dark soliton.
We also remark that all the results reduced from the CH2 system reproduce the corresponding ones obtained recently
by an analysis of the HS2 system [29].
\par
\bigskip
\noindent{\bf Remark 4.2.}\par
\noindent Under the scaling (4.16), the reciprocal transformation (2.1) and equations (2.2)-(2.5) remain the same form. The bilinear equations (2.11), (2.12) and (2.14) reduce respectively to
$$D_{\hat\tau}D_{\hat y}\hat f\cdot\hat f-{\kappa^2\over \hat \rho_0^2}\,(\hat f^2-\hat{\tilde g}\hat g)=0, \eqno(4.29)$$
$${\rm i}D_{\hat\tau}\hat{\tilde g}\cdot \hat g+\hat\rho_0\,(\hat f^2-\hat{\tilde g}\hat g)=0, \eqno(4.30)$$
$$D_{\hat\tau}D_{\hat y} \hat{\tilde g}\cdot \hat g-{\rm i}\,{\kappa^2\over \hat \rho_0^2}\,D_{\hat\tau} \hat{\tilde g}\cdot \hat g+{\rm i}\hat \rho_0\,D_{\hat y} \hat{\tilde g}\cdot \hat g=0, \eqno(4.31)$$
whereas the bilinear equation (2.13) reduces to (4.29) when coupled with (2.11).
\par 
\bigskip
\noindent{\it 4.3. Reduction to the HS equation} \par
\bigskip
\noindent The HS equation (1.5) can be reduced from either the short-wave limit of the CH equation or the scaling limit of the HS2 system, as shown in Figure 5.
The former reduction has been performed in [30]. To attain the latter reduction, we employ the same scaling variables as those given by (4.1) and
find that the resulting expressions reproduce those obtained in [30]. 
The reduction process can be established in parallel with that for the CH2 system, and hence
the detail of the computation is omitted here.  \par
The parametric representation of the $N$-soliton solution can be obtained by taking the scaling limit of (4.22), (4.23) and (4.25). 
It leads, after removing the hat appended to the variables for simplicity, to 
$$ u=-{2\over\kappa^2}\,({\rm ln}\,  f)_{\tau\tau}, \eqno(4.32a)$$
$$ x ={ y\over \kappa}-{2\over\kappa^2}\,({\rm ln}\,  f)_{\tau}+ d, \eqno(4.32b)$$
with
$$ f=\sum_{\mu=0,1}{\rm exp}\left[\sum_{j=1}^N\mu_j\xi_j
+\sum_{1\le j<l\le N}\mu_j\mu_l\gamma_{jl}\right], \eqno(4.33a)$$
$$\xi_j= k_j( y-c_j\tau- y_{j0}), 
\quad   c_j= -{2\kappa\over   k_j^2}, \ (j=1, 2, ..., N), \eqno(4.33b)$$  
$${\rm e}^{\gamma_{jl}}=\left({ k_j- k_l\over k_j+ k_l}\right)^2, \ (j, l=1, 2, ..., N; j\not=l). \eqno(4.33c)$$
 \par
 The one-soliton solution is given by
$$ u=-{2\over k^2}{1\over \cosh^2{\xi\over 2}}, \quad X=x-\tilde ct-x_0={\xi\over \kappa k}-{2\over \kappa k}\,\tanh\,{\xi\over 2}, \eqno(4.34a)$$
with
$$\xi=k(y-c\tau-y_0),\quad c=-{2\kappa\over k^2}, \quad \tilde c={c\over\kappa}. \eqno(4.34b)$$
 The above parametric solution takes the form of a cusp soliton. This can be confirmed simply by computing the derivative $u_X(=u_\xi/X_\xi)$ from (4.34), giving $u_X=4\kappa/(k\,\sinh\,\xi)$.
 Thus, $\lim_{X\rightarrow \pm 0}u_X=\pm\infty$, showing that the slope of the soliton becomes infinite at the crest.
\par
\bigskip
\noindent {\bf 5. Concluding remarks}\par
\bigskip
\noindent An intriguing feature of the CH equation is the existence of  peakons  which  mimic Stokes' limiting solitary waves in the classical
shallow water wave theory [31]. The peakons can be reduced from the smooth solitons by taking the zero dispersion limit $\kappa\rightarrow 0$. See, for example [32,  33].
Since the CH2 system under consideration is an integrable generalization of the CH equation, one can expect that it exhibits peakons as well.
The detailed analysis of the one-soliton solution (3.4) reveals that the peakon can not be produced from the smooth soliton in any limiting procedure.
On the other hand, another integrable CH2 system (2.31)  
admits peakons [6]. However, the general $N$-peakon solution  is  still unavailable for this system.  In addition,  whether peakons can be
reduced from smooth solitons or not has not been resolved.
The complete classification of traveling wave solutions of the CH2 system has not been performed yet for both periodic and nonperiodic
boundary conditions. Specifically, as for the existence of multi-valued solutions, no
decisive answer exists even today.
These interesting problems will be considered in a future work.
\par
\bigskip
\noindent {\bf Acknowledgements} \par
\bigskip
\noindent This work was partially supported by Yamaguchi University Foundation. 
The author appreciated critical review comments from two anonymous reviewers, which greatly improve an earlier draft of the manuscript.
\par
\bigskip
\noindent {\bf Appendix A. Proof of Proposition 2.2.}\par
\bigskip
\noindent  First, we show that the solutions of the bilinear equations (2.11) and (2.12) solve (2.7).  Upon substituting (2.9) and (2.10) into (2.7), the
equation to be proved becomes $P=0$, where
$$P\equiv \left\{{1\over\rho_0}+\left({\rm ln}\,{\tilde f\over f}\right)_y\right\}\left\{\rho_0-{\rm i}\left({\rm ln}\,{\tilde g\over g}\right)_\tau\right\}-1.$$
Invoking the definition of the bilinear operator (2.15), $P$ is rewritten in the form
$$P=\left\{\left({1\over\rho_0}\tilde f f+D_y\tilde f\cdot f\right)\left(\rho_0\tilde g g-{\rm i}D_\tau\tilde g\cdot g\right)-\tilde f\tilde g fg\right\}/(\tilde f\tilde g fg).$$
This expression becomes zero by virtue of (2.11) and (2.12). \par
To preceed, we introduce (2.9) and (2.10) into (2.8), and  obtain
$$\left\{\rho_0-{\rm i}\left({\rm ln}\,{\tilde g\over g}\right)_\tau\right\}\left\{{\kappa^2\over \rho_0^2}+{\rm i}\left({\rm ln}\,{\tilde g\over g}\right)_y\right\}
=\left({\rm ln}\,{\tilde f\over f}\right)_\tau\left\{{1\over\rho_0}+\left({\rm ln}\,{\tilde f\over f}\right)_y\right\}$$
$$-\left[\left\{\rho_0-{\rm i}\left({\rm ln}\,{\tilde g\over g}\right)_\tau\right\}\left({\rm ln}\,{\tilde f\over f}\right)_{\tau y}\right]_y+\kappa^2\left\{{1\over\rho_0}+\left({\rm ln}\,{\tilde f\over f}\right)_y\right\}.$$
In view of (2.11) and (2.12), the second term on the right-hand side of the above equation is modified as
$$\left[\left\{\rho_0-{\rm i}\left({\rm ln}\,{\tilde g\over g}\right)_\tau\right\}\left({\rm ln}\,{\tilde f\over f}\right)_{\tau y}\right]_y=\left({\rm ln}\,{\tilde g g\over \tilde f f}\right)_{\tau y}.$$
Inserting this relation  and using (2.11) and (2.12), the equation to be proved reduces to $Q=0$, where
$$Q\equiv \left({\rm ln}\,{\tilde g g\over \tilde f f}\right)_{\tau y}+{\rm i}\rho_0\,{\tilde f f\over \tilde g g}\left({\rm ln}\,{\tilde g\over g}\right)_y
-{1\over \rho_0}{\tilde g g\over \tilde f f}\left({\rm ln}\,{\tilde f\over f}\right)_\tau+{\kappa^2\over\rho_0}\left({\tilde f f\over \tilde g g}-{\tilde g g\over \tilde f f}\right).$$
It now follows from the the definition of the bilinear operators that
$$({\rm ln}\,\tilde f f)_{\tau y}={D_\tau D_y\tilde f\cdot f\over \tilde f f}-{1\over (\tilde f f)^2}(D_\tau\tilde f\cdot f)(D_y\tilde f\cdot f),$$
$$({\rm ln}\,\tilde g g)_{\tau y}={D_\tau D_y\tilde g\cdot g\over \tilde g g}-{1\over (\tilde g g)^2}(D_\tau\tilde g\cdot g)(D_y\tilde g\cdot g).$$
Substituting these identities into the first term of $Q$ and rewriting the second and third terms by means of the bilinear operators, $Q$ recasts to
$$Q={D_\tau D_y\tilde g\cdot g\over \tilde g g}+{\rm i}\,{D_y\tilde g\cdot g\over (\tilde g g)^2}\,({\rm i}D_\tau\tilde g\cdot g+ \rho_0\tilde f f)
-{D_\tau D_y\tilde f\cdot f\over \tilde f f}$$
$$+{D_\tau\tilde f\cdot f\over (\tilde f f)^2}\left(D_y\tilde f\cdot f-{1\over\rho_0}\,\tilde g g\right)+{\kappa^2\over\rho_0}\left({\tilde f f\over \tilde g g}-{\tilde g g\over \tilde f f}\right).$$
This expression turns out to be zero by virtue of (2.11)-(2.14). \par
\bigskip
\leftline{\bf Appendix B. Proof of Theorem 2.2} \par
\bigskip
\noindent
In this appendix, we show that the tau-functions (2.21) and (2.22) solve the system of bilinear equations (2.11)-(2.14).
We use a mathematical induction similar to that has been employed for the proof of the $N$-soliton solution of the nonlinear network equations [34].
Since the proof can be performed  in a similar manner for all equations, we describe the proof of (2.11) in some detail, and outline the proof for
other three equations. \par
First, we substitute the tau-functions $f$ and $\tilde f$ from (2.21) into the bilinear equation (2.11) and use the formula
$$D_\tau^mD_y^n\,{\rm exp}\left[\sum_{i=1}^N\mu_i\xi_i\right]\cdot {\rm exp}\left[\sum_{i=1}^N\nu_i \xi_i\right]$$
$$=\left\{-\sum_{i=1}^N(\mu_i-\nu_i)k_i c_i\right\}^m\left\{\sum_{i=1}^N(\mu_i-\nu_i)k_i\right\}^n{\rm exp}\left[\sum_{i=1}^N(\mu_i+\nu_i)\xi_i\right],\ (m, n=0, 1, 2, ...), \eqno(B. 1) $$
to show that the equation to be proved becomes
 $$\sum_{\mu, \nu=0,1}\left[\left\{\sum_{i=1}^N(\mu_i-\nu_i)k_i+{1\over \rho_0}\right\}\,{\rm exp}\left[-\sum_{i=1}^N(\mu_i-\nu_i)\phi_i\right]
 -{1\over \rho_0}\,{\rm exp}\left[-{\rm i}\sum_{i=1}^N(\mu_i-\nu_i)\psi_i\right]\right]$$
 $$\times {\rm exp}\left[\sum_{i=1}^N(\mu_i+\nu_i)\xi_i+\sum_{1\leq i<j\leq N}(\mu_i\mu_j+\nu_i\nu_j)\gamma_{ij}\right]=0. \eqno(B. 2)$$
 \par
Let $P_{m, n}$ be the coefficient of the factor ${\rm exp}\left[\sum_{i=1}^n \xi_i+\sum_{i=n+1}^m 2\xi_i\right]\ (1\leq n<m\leq N)$ on the left-hand side of (B.2). 
This coefficient is obtained if one performs the
summation with respect to $\mu_i$ and $\nu_i$  under the conditions
$\mu_i+\nu_i=1\ (i=1, 2, ..., n), \ \mu_i=\nu_i=1\ (i=n+1, n+2, ..., m),
\ \mu_i=\nu_i=0\ (i=m+1, m+2, ..., N).$
We then introduce the new summation indices $\sigma_i$ by the relations $\mu_i=(1+\sigma_i)/2,\ \nu_i=(1-\sigma_i)/2$ for $i=1, 2, ..., n$, where $\sigma_i$ takes either the value $+1$ or $-1$,
so that  $\mu_i\mu_j+\nu_i\nu_j=(1+\sigma_i\sigma_j)/2$. \par
Consequently,  $P_{m, n}$ can be rewritten in the form
$$P_{m,n}=\sum_{\sigma=\pm 1}\left[\left\{\sum_{i=1}^n\sigma_ik_i+{1\over \rho_0}\right\}{\rm exp}\left[-\sum_{i=1}^n\sigma_i\phi_i\right]
-{1\over \rho_0}\,{\rm exp}\left[-{\rm i}\sum_{i=1}^n\sigma_i\psi_i\right]\right]$$
$$\times {\rm exp}\left[{1\over 2}\sum_{1\leq i<j\leq n}(1+\sigma_i\sigma_j)\gamma_{ij}+\sum_{i=1}^m\sum_{\substack{j=n+1\\ (j\not=i)}}^m\gamma_{ij}\right]. \eqno(B. 3)$$
If we invoke (2.24) and  (2.28)-(2.30) as well as the definition of $\sigma_i$, we deduce
$${\rm exp}\left[-\sum_{i=1}^n \sigma_i\phi_i\right]=\prod_{i=1}^n\left[{{\rm sgn}\,c_i\over \sqrt{\rho_0^2+\kappa^4}}{d_i-\kappa^2\rho_0\sigma_ik_i\over 1+\rho_0\sigma_ik_i}\right], \eqno(B. 4)$$   
$$ {\rm exp}\left[-{\rm i}\sum_{i=1}^n \sigma_i\psi_i\right]=\prod_{i=1}^n\left[{{\rm sgn}\,c_i(d_i-{\rm i}\rho_0^2\sigma_ik_i)\over \sqrt{\rho_0^2+\kappa^4}}\right], \eqno(B. 5)$$
$${\rm exp}\left[{1\over 2}\sum_{1\leq i<j\leq n}(1+\sigma_i\sigma_j)\gamma_{ij}\right]
=\prod_{1\leq i<j\leq n}\left[{(d_i-d_j)^2+\rho_0^4(\sigma_ik_i-\sigma_jk_j)^2\over (d_i-d_j)^2+\rho_0^4(\sigma_ik_i+\sigma_jk_j)^2}\right]. \eqno(B. 6)$$
Substituting (B. 4)-(B. 6) into (B. 3), $P_{m, n}$ becomes
$$P_{m, n}=c_{m, n}\sum_{\sigma=\pm 1}\left[\left(\sum_{i=1}^n\rho_0\sigma_ik_i+1\right)\prod_{i=1}^n{d_i-\kappa^2\rho_0\sigma_ik_i\over 1+\rho_0\sigma_ik_i}
 -\prod_{i=1}^n(d_i-{\rm i}\rho_0^2\sigma_ik_i)\right]$$
 $$\times \prod_{1\leq i<j\leq n}\left[(d_i-d_j)^2+\rho_0^4(\sigma_ik_i-\sigma_jk_j)^2\right], \eqno(B. 7)$$
 where $c_{m, n}$ is a multiplicative factor independent of the summation indices $\sigma_i\ (i=1, 2, ..., n)$.
 To put (B. 7) into a more tractable form, we  introduce the new variables $r$ and $\theta_i$  by
$d_i+{\rm i}\rho_0^2k_i=r{\rm e}^{{\rm i}\theta_i}=rz_i$, where $z_i={\rm e}^{{\rm i}\theta_i}, r=\sqrt{d_i^2+\rho_0^4k_i^2}=\sqrt{\kappa^4+\rho_0^2}$.
Note that $r$ is a constant independent of $k_i$. To proceed, we substitute  the relation
$${d_i-\kappa^2\rho_0\sigma_ik_i\over 1+\rho_0\sigma_ik_i}={-\kappa^2d_i+\kappa^4+\rho_0^2-\rho_0^3\sigma_ik_i\over d_i-\kappa^2}, \eqno(B. 8)$$
which follows from (2.24) into the first term on the right-hand side of (B. 7) and then rewrite  $P_{m, n}$ in terms of the new variables $z_i$. Dropping a factor
independent of the summation indices $\sigma_i$, the equation  to be proved reduces to the following algebraic identity  in $z_1, z_2, ..., z_n$: 
\par
$P_n(z_1, z_2, ..., z_n)$
$$\equiv \sum_{\sigma=\pm 1}\Biggl[\left\{{r\over 2{\rm i}\rho_0}\sum_{i=1}^n\left(z_i^{\sigma_i}-z_i^{-\sigma_i}\right)+1\right\}
\prod_{j=1}^n\left\{-{\kappa^2\over 2r}\left(z_j+z_j^{-1}\right)+1-{\rho_0\over 2{\rm i}r}\left(z_j^{\sigma_j}-z_j^{-\sigma_j}\right)\right\}$$
$$-\prod_{j=1}^n\left\{{1\over 2}\left(z_j+z_j^{-1}\right)-{\kappa^2\over r}\right\}z_j^{-\sigma_j}\Biggr]
\prod_{1\leq i<j\leq n}\left(z_i^{\sigma_i}-z_j^{\sigma_j}\right)\left(z_i^{-\sigma_i}-z_j^{-\sigma_j}\right)=0,\ (n= 1, 2, ..., N).  \eqno(B. 9)$$
\par
The proof proceeds by mathematical induction.  The identity (B. 9) can be confirmed for $n=1, 2$ by a direct computation. Assume that $P_{n-2}=P_{n-1}=0$.
Then,
$$P_n|_{z_1=1}=2\left(1-{\kappa^2\over r}\right)\prod_{i=2}^n(1-z_i)(1-z_i^{-1})P_{n-1}(z_2, z_3, ..., z_n)=0. \eqno(B. 10)$$
$$P_n|_{z_1=z_2}=-2(z_1-z_1^{-1})^2\left\{{1\over 2}(z_1+z_1^{-1})-{\kappa^2\over r}\right\}^2$$
$$\times\prod_{j=3}^n(z_1-z_j)(z_1^{-1}-z_j^{-1})(z_1^{-1}-z_j)(z_1-z_j^{-1})P_{n-2}(z_3, z_4, ..., z_n)=0. \eqno(B. 11)$$
The function $P_n$ is symmetric with respect to $z_1, z_2, ..., z_n$ and  invariant under the transformation $z_i \rightarrow z_i^{-1}$ for arbitrary $i$.
When coupled with the above two properties (B. 10) and (B. 11), one can see that $P_n$ is factored by a function
$$\prod_{i=1}^n(z_i-1)(z_i^{-1}-1)\prod_{1\leq i<j\leq n}(z_i-z_j)(z_i-z_j^{-1})(z_i^{-1}-z_j)(z_i^{-1}-z_j^{-1}). \eqno(B. 12)$$
  It turns out from this expression that
$$\prod_{i=1}^nz_i^2\prod_{1\leq i<j\leq n}(z_iz_j)^2P_n=A_n\prod_{i=1}^n(z_i-1)(1-z_i)\prod_{1\leq i<j\leq n}(z_i-z_j)^2(z_iz_j-1)^2, \eqno(B. 13)$$
where $A_n$ is a polynomial  of $z_1, z_2, ..., z_n$.
 The left-hand side of (B. 13) is a polynomial whose degree in $z_1, z_2, ..., z_n$ is at most $2n^2+2n$ whereas that of the right-hand side is
 $3n^2-n$ at least. This is impossible for $n\geq 4$ except $P_n\equiv 0$. 
 The identity $P_3=0$ can be checked by a direct computation, implying that the identity (B. 9) holds for all $n$.  \par
 The bilinear equations (2.12), (2.13) and (2.14) reduce, after substituting the tau-functions (2.21) and (2.22), to the algebraic identities
 $Q_n=0, R_n=0$ and $S_n=0$, respectively, where \par
 \bigskip
 $Q_n(z_1, z_2, ..., z_n)$
 $$\equiv \sum_{\sigma=\pm 1}\Biggl[\biggl[\prod_{i=1}^n\left\{{1\over 2}(z_i+z_i^{-1})-{\kappa^2\over r}\right\}+{1\over 2}\sum_{i=1}^n\left(z_i^{\sigma_i}-z_i^{-\sigma_i}\right)
 \prod_{\substack{j=1\\ (j\not=i)}}^n\left\{{1\over 2}(z_j+z_j^{-1})-{\kappa^2\over r}\right\}\biggr]\prod_{j=1}^nz_j^{-\sigma_j}$$
$$-\prod_{i=1}^n\left\{-{\kappa^2\over 2r}(z_i+z_i^{-1})+1-{\rho_0\over 2{\rm i}r}\left(z_i^{\sigma_i}-z_i^{-\sigma_i}\right)\right\}\Biggr]
\prod_{1\leq i<j\leq n}\left(z_i^{\sigma_i}-z_j^{\sigma_j}\right)\left(z_i^{-\sigma_i}-z_j^{-\sigma_j}\right),$$
$$ (n=1, 2, ..., N).  \eqno(B. 14)$$
\par
 $R_n(z_1, z_2, ..., z_n)$
 $$\equiv \sum_{\sigma=\pm 1}\Biggl[\left\{1+{r\over 2{\rm i}\rho_0}\sum_{i=1}^n\left(z_i^{\sigma_i}-z_i^{-\sigma_i}\right)\right\}\biggl[\sum_{i=1}^n\left(z_i^{\sigma_i}-z_i^{-\sigma_i}\right)
\prod_{\substack{j=1\\ (j\not=i)}}^n\left\{{1\over 2}(z_j+z_j^{-1})-{\kappa^2\over r}\right\}\biggr]$$
$$-{\kappa^2r\over \rho_0^2}\sum_{i=1}^n\left(z_i^{\sigma_i}-z_i^{-\sigma_i}\right)\prod_{i=1}^n\left\{{1\over 2}(z_i+z_i^{-1})-{\kappa^2\over r}\right\}\Biggr]$$
$$\times\prod_{i=1}^n\left\{-{\kappa^2\over 2r}(z_i+z_i^{-1})+1-{\rho_0\over 2{\rm i}r}\left(z_i^{\sigma_i}-z_i^{-\sigma_i}\right)\right\}$$
$$\times \prod_{1\leq i<j\leq n}\left(z_i^{\sigma_i}-z_j^{\sigma_j}\right)\left(z_i^{-\sigma_i}-z_j^{-\sigma_j}\right), \quad  (n=1, 2, ..., N).  \eqno(B. 15)$$
\par
 $S_n(z_1, z_2, ..., z_n)$
 $$\equiv \sum_{\sigma=\pm 1}\Biggl[\left\{{\kappa^2\over r}+{1\over 2}\sum_{i=1}^n\left(z_i^{\sigma_i}-z_i^{-\sigma_i}\right)\right\}\sum_{i=1}^n\left(z_i^{\sigma_i}-z_i^{-\sigma_i}\right)
\prod_{\substack{j=1\\ (j\not=i)}}^n\left\{{1\over 2}(z_j+z_j^{-1})-{\kappa^2\over r}\right\}\biggr]$$
$$+\sum_{i=1}^n\left(z_i^{\sigma_i}-z_i^{-\sigma_i}\right)\prod_{i=1}^n\left\{{1\over 2}(z_i+z_i^{-1})-{\kappa^2\over r}\right\}\Biggr]$$
$$\times \prod_{i=1}^nz_i^{-\sigma_i}\prod_{1\leq i<j\leq n}\left(z_i^{\sigma_i}-z_j^{\sigma_j}\right)\left(z_i^{-\sigma_i}-z_j^{-\sigma_j}\right), \quad  (n=1, 2, ..., N).  \eqno(B. 16)$$
The polynomials $Q_n, R_n$ and $S_n$ are shown to be  factored by a function (B. 12). 
Applying the similar induction argument to that used in proving (B. 9), one can establish the identities
$Q_n=0, R_n=0$ and $S_n=0$. This completes the proof of theorem 2.2. \par

\newpage
\leftline{\bf Reference} \par
\baselineskip=5.5mm
\begin{enumerate}[{[1]}]
\item Olver P and Rosenau P 1996 Tri-Hamiltonian duality between solitons and solitary-wave solutions having compact support {\it Phys. Rev. E} {\bf 53} 1900-6
\item Zakharov V E 1980  The inverse scattering method {\it Solitons ( Topics in Current Physics {\bf vol 17})} 
              ed R K Bullough and D J Caudrey (New York: Springer) pp 243-85
\item Ito M 1982  Symmetries and conservation laws of a coupled nonlinear wave equation {\it Phys. Lett. A} {\bf 91} 335-8
\item Liu S-Q and Zhang Y 2005  Deformation of semisimple bihamiltonian structures of hydrodynamic type {\it J. Geom. Phys.} {\bf 54} 427-53
\item Falqui G 2006 On a Camassa-Holm type equation with two dependent variables {\it J. Phys. A: Mth. Gen.} {\bf 39} 327-42
\item Chen M, Liu S-Q and Zhang Y 2006 A two-component generalization of the Camassa-Holm equation and its solutions {\it Lett. Math. Phys.} {\bf 75} 1-15
\item Constantin A and Ivanov R I 2008 On an integrable two-component Camassa-Holm shallow water system {\it Phys. Lett.} {\bf A 372} 7129-32
\item Ivanov R I 2009 Two-component integrable systems modelling shallow water waves: The constant vorticity case {\it Wave Motion} {\bf 46} 389-96
\item Holm D D and Ivanov R I 2010 Multi-component generalization of the CH equation: geometric aspects, peakons and numerical examples {\it J. Phys. A: Math. Theor.} {\bf 43} 492001
\item Ablowitz M J and Segur H 1981 {\it Solitons and the Inverse Scattering Transform (SIAM Studies in Applied Mathematics {\rm vol 4})} (Philadelphia, PA: SIAM)
\item Ablowitz M J and Clarkson P A 1991 {\it Nonlinear Evolution Equations and Inverse Scattering (London Mathematical Society Lecture Notes Series \#149)} (Cambridge: Cambridge University Press)
\item Escher J, Lechtenfeld O and Yin Z 2007 Well-posedness and blow-up phenomena for the 2-component Camassa-Holm equation {\it Discrete Continuous Dyn. Syst.} {\bf 19} 493-513
\item Gui G and Liu Y 2010  On the global existence and wave-breaking criteria for the two-component Camassa-Holm system {\it J. Funct. Anal.} {\bf 258} 4251-78
\item Li J B and Li Y S 2008  Bifurcations of travelling wave solutions for a two-component Camassa-Holm equation {\it Acta Math. Sinica} {\bf 24} 1319-30
\item Dutykh D and Ionescu-Kruse D 2016 Travelling wave solutions for some two-component shallow water models {\it J. Diff. Eqs.} {\bf 261} 1099-114
\item Holm D D and Ivanov R I 2011 Two-component CH system: inverse scattering, peakons and geometry {\it Inverse  Problems} {\bf 27} 045013
\item  Camassa R and Holm D D 1993 An integrable shallow water equation with peaked solitons {\it Phys. Rev. Lett.} {\bf 71} 1661-4
\item Camassa R, Holm D and Hyman J 1994 A new integrable shallow water wave equation {\it Adv. Appl. Mech.} {\bf 31} 1-33
\item Holm D D and Ivanov R I 2010 Smooth and peaked solitons of the CH equation {\it J. Phys. A: Math. Theor.} {\bf 43} 434003
\item Hunter J and Saxton R 1991  Dynamics of director fields {\it SIAM J. Appl. Math.} {\bf 51} 1498-521
\item Matsuno Y 2005 Parametric representation for the multisoliton solution of the Camassa-Holm equation {\it J. Phys. Soc. Jpn.} {\bf 74} 1983-87
\item  Matsuno Y 2013 B\"acklund transformation and smooth multisoliton solutions for a modified Camassa-Holm equation with cubic nonlinearity {\it J. Math. Phys.} {\bf 54} 051504
\item Matsuno Y 2014 Smooth and singular multisoliton solutions of a modified Camassa-Holm equation wih cubic nonlinearity and linear dispersion {\it J. Phys. A: Math. Theor.} {\bf 47} 125203
\item  Hirota R 1980  Direct Methods in Soliton Theory {\it Solitons (Topics in Current Physics {\bf vol 17})}
        ed RK Bullough and DJ Caudrey  (New York: Springer) pp 157-76
\item  Matsuno Y 1984 {\it Bilinear Transformation Method} (New York: Academic)
\item Date E, Jimbo M and Miwa T 1983 Method for generating discrete soliton equations. V {J. Phys. Soc. Jpn.} {\bf 52} 766-71
\item Jimbo M and Miwa T 1983 Solitons and infinite dimensional Lie algebras {\it Publ. RIMS Kyoto Univ.} {\bf 19} 943-1001
\item Wu C Z 2006 On solutions of the two-component Camassa-Holm system {\it J. Math. Phys} {\bf 47} 083513
\item  Lau S, Feng B F and Yao R 2016 Multi-soliton solution to the two-component Hunter-Saxton equation {\it Wave Motion} {\bf 65} 17-28
\item Matsuno Y, 2006  Cusp and loop soliton solutions of short-wave models for the Camassa-Holm and Degasperis-Procesi equations {\it Phys. Lett. A} {\bf 359} 451-457
\item Whitham G B 1974 {\it Linear and Nonlinear Waves} (New York: John Wiley \& Sons)
\item Parker A and Matsuno Y 2006 The peakon limits of soliton solutions of the Camassa-Holm equation {\it J. Phys. Soc. Jpn.} {\bf 75} 124001
\item Matsuno Y 2007 The peakon limit of the $N$-soliton solution of the Camassa-Holm equation {\it J. Phys. Soc. Jpn.} {\bf 76} 034003
\item  Hirota R 1973 Exact $N$-soliton solution of nonlinear lumped self-dual network equations {\it J. Phys. Soc. Jpn.} {\bf 35} 289-294

\end{enumerate}

\end{document}